\documentclass[superscriptaddress,pre,twocolumn,nofootinbib]{revtex4}

\usepackage{amsmath}
\usepackage{graphicx}
\usepackage{color}
\newcommand{\be}{\begin{equation}}
\newcommand{\ee}{\end{equation}}
\newcommand{\bea}{\begin{eqnarray}}
\newcommand{\eea}{\end{eqnarray}}

\begin{document}
\sloppy


\title{The Boltzmann-Poisson equation with a central body: analytical \\
solutions in one and two dimensions}

\author{Pierre-Henri Chavanis}
\email{chavanis@irsamc.ups-tlse.fr}
\affiliation{Laboratoire de Physique Th\'eorique, Universit\'e de Toulouse,
CNRS, UPS, France}

\begin{abstract}
We consider an isothermal self-gravitating system surrounding a central body. This model can represent a galaxy or a globular cluster harboring a central black hole.  It can also represent  a gaseous atmosphere surrounding a protoplanet. In three dimensions, the Boltzmann-Poisson equation must be solved  numerically in order to obtain the density profile of the gas 
[Chavanis {\it et al.}, Phys. Rev. E {\bf 109}, 014118 (2024)]. In one and two dimensions, we show that the Boltzmann-Poisson equation can be solved analytically. We obtain explicit analytical expressions of the density profile around a central body which generalize the analytical solutions found by Camm (1950) and Ostriker (1964) in the absence of a central body. Our results also have applications for self-gravitating Brownian particles (Smoluchowski-Poisson system), for the chemotaxis of bacterial populations (Keller-Segel model), and for two-dimensional point vortices (Onsager's model). 
In the case of bacterial populations, the central body could represent a supply of
``food'' that attracts the bacteria (chemoattractant). In the case of
two-dimensional vortices, the central body could be a central vortex.
 \end{abstract}

\maketitle

\section{Introduction}

In the theory of stars and galaxies, a special attention has been given to
isothermal configurations \cite{emden,chandrass,paddy,dvs1,dvs2,found,ijmpb}.
This is partly due to their mathematical simplicity and also because an
isothermal (Boltzmann) distribution corresponds to the equilibrium state
predicted by statistical mechanics. It is however well-known that no strict
statistical equilibrium state exists for self-gravitating systems because these
systems have the tendency to evaporate \cite{amba,spitzer1940,chandra}. In
particular, an isothermal self-gravitating system is unphysical because it has
an infinite mass. A possibility to overcome this problem is to use a truncated
Boltzmann distribution like the Michie-King \cite{michie,king} model which takes
into account the evaporation of high energy particles. This model has a finite
mass. It differs from a pure isothermal distribution in that the velocity
dispersion is position-dependent. However, some regions of stars and galaxies
can be approximated by an isothermal distribution with a uniform velocity
dispersion, so this distribution is physical at least locally. One can also
study the statistical mechanics of a self-gravitating system enclosed within a
box in order to avoid the infinite mass problem and make the analysis well-posed
mathematically. This academic study led to the discovery of important physical
concepts such as the existence of negative specific heat, the inequivalence of statistical ensembles, and the gravothermal catastrophe  \cite{antonov,lbw}. These concepts remain valid for realistic models of globular clusters described by
the Michie-King distribution \cite{katzking,clm1}. During their secular
(collisional) evolution,  globular clusters follow the King sequence. As their central density increases due to evaporation, they may become thermodynamically
unstable and experience core collapse  leading to a finite time singularity
\cite{larson,hachisu,lbe,cohn}. A binary star \cite{henonbinary} is
formed in the post-collapse regime and
releases so much energy that the halo re-expands in a
self-similar manner~\cite{inagakilb,hs}. Finally, a series of gravothermal oscillations follows~\cite{sugimoto}.

On the mathematical point of view, self-gravitating isothermal configurations are obtained by solving the Boltzmann-Poisson equation. Unfortunately, no analytical solution is known in $d=3$ dimensions and the equation must be solved numerically \cite{emden,chandra}. However, analytical solutions have been found in $d=1$ and $d=2$ dimensions by Camm \cite{camm} and Ostriker \cite{ostriker}, and rediscovered by various authors (a detailed list of references is given at the beginning of Secs. \ref{sec.analytique1d} and \ref{sec.analytique2d}). These solutions describe self-gravitating isothermal sheets or filaments (rods). They also provide explicit self-consistent models of self-gravitating systems that can be valuable for theoretical works (e.g., for stability analysis). One interest of these models in $d=1$ and $d=2$ dimensions is that the presence of an artificial box is not necessary to make the mass finite, i.e., a statistical equilibrium state can exist in an unbounded domain. 

In many astrophysical systems, the constituents (elementary particles, atoms,
molecules, stars...) are in gravitational interaction but they are also
subject to the attraction of a central body. This can be a protoplanet
surrounded by a gaseous atmosphere or a black hole at the center of a globular
cluster or at the center of a galaxy. It is therefore interesting to study the
configuration of an isothermal self-gravitating system surrounding a central
body. Here again, the most relevant three dimensional case must be solved
numerically.  Recently,  we have considered this problem from the viewpoint of
thermodynamics and statistical mechanics and we have described interesting phase
transitions between gaseous and condensed configurations exhibiting a cusp 
at the contact with the central body \cite{css}.    In the present paper, as a
complement, we provide analytical solutions of the Boltzmann-Poisson
equations with a central body in one and two dimensions that generalize the analytical solutions
found by Camm \cite{camm} and Ostriker \cite{ostriker} in the absence of a central
body. These solutions may be useful to better understand the structure of an
isothermal self-gravitating atmosphere around a central body.

The paper is organized as follows. In Sec. \ref{sec_fund}, we
recall the basic equations describing the equilibrium state of a
self-gravitating isothermal gas surrounding a central body.
In Secs. \ref{sec.analytique1d} and \ref{sec.analytique2d}, we analytically
solve the differential equation for the mass profile of the gas in one and two
dimensions in a finite and infinite domain. In Appendix \ref{sec_sgbp}, we
present the equations describing the dynamical evolution of a gas of
self-gravitating Brownian particles in the presence of a central body and
establish the virial theorem. In Appendix \ref{sec_dye}, we summarize the
results obtained in previous works concerning the dynamical evolution of
self-gravitating Brownian particles in different dimensions of space. In
Appendix \ref{sec_a2d}, we discuss the analogy between self-gravitating systems
and two-dimensional vortices. Finally, in Appendices
\ref{app.boltzpoisson1d} and \ref{app.boltzpoisson2d}, we analytically solve the
Boltzmann-Poisson equation with a central body in one and two dimensions and
recover the results of the main text in a different manner.

\section{Fundamental equations for a self-gravitating system in the presence of a central body}
\label{sec_fund}

We consider an isothermal self-gravitating atmosphere of mass $M=Nm$ (where $N$ is the number of constituents of mass $m$) and temperature $T$ surrounding a central body of mass $M_*$ and radius $R_*$. We assume that both the atmosphere and the central body are spherically symmetric. In this section, we consider a system in $d$ dimensions.

\subsection{Newton's law}

In the region of the gas surrounding the central body ($r\ge R_*$),  the
gravitational potential $\Phi$ produced by the gas satisfies the Poisson
equation
\be
\Delta\Phi=S_{d}G\rho,
\label{poisson}
\ee
where $\rho$ is the mass density of the gas, $G$ is the gravity constant and $S_d=2\pi^{d/2}/\Gamma(d/2)$ is the surface of a
unit sphere in $d$ dimensions. In particular, $S_1=2$, $S_2=2\pi$ and $S_3=4\pi$.  On the other hand, the gravitational potential produced
by the central body satisfies the Laplace equation
\be
\Delta\Phi_{\rm ext}=0.
\label{laplace}
\ee
If we introduce the total gravitational potential
\be
\Phi_{\rm tot}=\Phi+\Phi_{\rm ext},
\label{min}
\ee
we get
\be
\label{feel}
\Delta\Phi_{\rm tot}=S_{d}G\rho.
\ee
For a spherically symmetric system, the Poisson equation (\ref{feel}) can be written as
\be
\frac{1}{r^{d-1}}\frac{d}{dr}\left (r^{d-1}\frac{d\Phi_{\rm tot}}{dr}\right
)=S_{d}G\rho.
\ee
Integrating this equation between $R_*$ and $r$, we obtain
\be
r^{d-1}\frac{d\Phi_{\rm tot}}{dr}-R_*^{d-1}\frac{d\Phi_{\rm tot}}{dr}(R_*)=G
M(r),
\ee
where
\be
\label{mr}
M(r)=\int_{R_*}^{r}\rho(r')S_d {r'}^{d-1}\, dr'
\ee
denotes the mass of gas contained between the spheres of radius $R_*$ and
$r$; it satisfies $M(R_{*})=0$ and $M(R)=M$ in a finite domain or $M(+\infty)=M$ in an infinite domain (in $d=1$ or $d=2$ dimensions). For future reference, we
note
that
\be
\label{eq.rhom'}
\rho(r)=\frac{M'(r)}{S_{d} r^{d-1}}.
\ee
Using [see Eq. (28) of \cite{css}]
\be
\label{mado}
\frac{d\Phi_{\rm
tot}}{dr}(R_*)=\frac{d\Phi_{\rm ext}}{dr}(R_*)=\frac{GM_{*}}{R_*^{d-1}},
\ee
which is the consequence of the presence of the central body, we finally obtain the $d$-dimensional Newton law
in the presence of a central body
\be
\label{oas}
\frac{d\Phi_{\rm tot}}{dr}=\frac{G(M_{*}+M(r))}{r^{d-1}}.
\ee
This equation can be directly
obtained from Eq. (D3) of \cite{css} with the substitution $\Phi(r)\rightarrow
\Phi_{\rm tot}(r)$ and $M(r)\rightarrow M_{\rm tot}(r)=M_*+M(r)$. It can also be obtained by integrating individually Eqs. (\ref{poisson}) and (\ref{laplace}) and summing the resulting expressions.

If the system is contained within a box of radius $R$, then
for $r\ge R$, we have 
\be
\frac{d\Phi_{\rm tot}}{dr}=\frac{G (M_*+M)}{r^{d-1}}.
\ee
For $r\ge R$ the total gravitational potential  is given by 
\be
\Phi_{\rm tot}(r)=-\frac{1}{d-2}\frac{G(M_*+M)}{r^{d-2}}\quad (d\neq 2),
\ee
\be
\Phi_{\rm tot}(r)=G(M_*+M)\ln \left (\frac{r}{R}\right )\quad (d= 2).
\ee
In particular,
\be
\Phi_{\rm tot}(R)=-\frac{1}{d-2}\frac{G(M_*+M)}{R^{d-2}}\quad (d\neq 2),
\ee
\be
\Phi_{\rm tot}(R)=0\quad (d= 2).
\ee

\subsection{Boltzmann-Poisson equation}

If we consider a gas at statistical equilibrium \cite{css}, the distribution function $f({\bf r},{\bf v})$ is given by the mean field Maxwell-Boltzmann distribution
\begin{equation}
f=A'e^{-\beta m (\frac{v^2}{2}+\Phi_{\rm tot})},
\label{adel}
\end{equation}
where $\beta=1/(k_BT)$ is the inverse temperature and $m$ the mass of a particle. Integrating this distribution over the velocity, we obtain the mean field Boltzmann distribution
\be
\rho=A e^{-\beta m\Phi_{\rm tot}}
\label{dion}
\ee
with $A=(2\pi/\beta m)^{d/2}A'$.
The normalization constants $A$ and $A'$ can be related to the total mass of gas $M=\int\rho\, d{\bf r}$.
Substituting the isothermal distribution (\ref{dion})  into the Poisson equation
(\ref{feel}), we
get the 
Boltzmann-Poisson equation
\be
\label{bpo}
\Delta\Phi_{\rm tot}=S_{d}GAe^{-\beta m\Phi_{\rm tot}}.
\ee
It can be shown that the
maximum entropy state for a non-rotating self-gravitating system is spherically
symmetric \cite{antonov} (see also footnote 11 in \cite{css}).
In that case, the Boltzmann-Poisson equation becomes
\be
\label{gim}
\frac{1}{r^{d-1}}\frac{d}{dr}\left (r^{d-1}\frac{d\Phi_{\rm tot}}{dr}\right
)=S_{d}GAe^{-\beta m\Phi_{\rm tot}}.
\ee
It has to be solved with the boundary condition (\ref{mado}). The
Boltzmann-Poisson equation (\ref{gim}) is equivalent to the  integrodifferential
equation [see Eq. (\ref{oas})]
\begin{equation}
\label{mid}
r^{d-1} \frac{d \Phi_{\rm tot}}{d r} = GM_*
+  G A \int_{R_*}^r  e^{-\beta m \Phi_{\rm tot}(r')} S_d\, {r'}^{d-1} \,dr'.
\end{equation}
On the other hand, the Boltzmann distribution (\ref{dion}) can be written as
\be
\rho=\rho_{0}e^{-\beta m (\Phi_{\rm tot}-\Phi_{\rm tot,0})},
\ee
where $\rho_0=\rho(R_{*})$ and $\Phi_{\rm tot,0}=\Phi_{\rm tot}(R_{*})$ are the
values of the density of the gas and of the total gravitational potential at the
contact with the central body. By an abuse of
langage, we will call them the central density and the central potential. Introducing
the dimensionless quantities 
\be
\psi=\beta m(\Phi_{\rm tot}-\Phi_{\rm tot,0})
\ee
and
\be
\xi=(S_d \beta
Gm\rho_{0})^{1/2}r,
\ee
the density (\ref{dion}) can be written as
\be
\label{rhopsi}
\rho=\rho_0e^{-\psi}
\ee
and the Boltzmann-Poisson equation (\ref{gim}) takes the form
\be
\label{emden}
\frac{1}{\xi^{d-1}}\frac{d}{d\xi}\left (\xi^{d-1}\frac{d\psi}{d\xi}\right )=e^{-\psi}.
\ee
This is the so-called Emden equation. The boundary conditions are
\be
\psi(\xi_{0})=0, \qquad \psi'(\xi_0)=\frac{\eta_0}{\xi_0},
\label{emdenbc}\ee
where $\xi_0$ and $\eta_0$ are defined by
\be
\xi_{0}=(S_d \beta Gm\rho_{0})^{1/2}R_{*},\qquad \eta_0=\frac{\beta
GmM_*}{R_{*}^{d-2}}.
\ee
We note that
\be
\label{piano}
\frac{\eta_0}{\xi_{0}}=\left (\frac{\beta G m}{S_d \rho_{0}}\right )^{1/2}\frac{M_{*}}{R_*^{d-1}},\qquad \frac{\xi}{\xi_0}=\frac{r}{R_*}.
\ee
Similarly to Eq. (\ref{mid}) the Emden equation (\ref{emden}) can be written in the form of an
integrodifferential equation
\be
\label{eq.equationintegrodiffpsicc}
\xi^{d-1}\frac{d \psi}{d \xi} = \eta_0 \xi_0^{d-2}
+ \int_{\xi_0}^\xi  e^{-\psi(\xi')} {\xi'}^{d-1}\,d\xi'.
\ee
This is equivalent to Newton's law [see Eq. (\ref{oas})]
\be
\label{mxi}
\frac{d\psi}{d\xi}=\eta_0\xi_0^{d-2}\left\lbrack
1+\frac{M(\xi)}{M_*}\right\rbrack \frac{1}{\xi^{d-1}}
\ee
with
\be
\label{mxib}
M(\xi)=\frac{M_*}{\eta_0\xi_0^{d-2}}\int_{\xi_0}^\xi  e^{-\psi(\xi')} {\xi'}^{d-1}\,d\xi'.
\ee

\subsection{The fundamental equation of hydrostatic equilibrium}
\label{sec_fy}

We can obtain the spatial structure of a self-gravitating gas
surrounding a
central
body in a different (but equivalent) manner by starting directly from the
condition of
hydrostatic equilibrium for $r\ge R_{*}$:
\be
\nabla P+\rho\nabla\Phi_{\rm tot}={\bf 0}.
\label{adel3}
\ee
For a spherically symmetric system,
using the Newton law (\ref{oas}), it can be rewritten as
\be
\frac{dP}{dr}=-\rho \frac{d\Phi_{\rm
tot}}{dr}=-\rho\frac{G(M_{*}+M(r))}{r^{d-1}}.
\label{vio}
\ee
Multiplying this equation by $r^{d-1}/\rho$, taking the derivative 
with respect to $r$ and using Eq. (\ref{eq.rhom'}),  we obtain
\be
\frac{1}{r^{d-1}}\frac{d}{dr}\left (\frac{r^{d-1}}{\rho}\frac{dP}{dr}\right
)=-S_{d}G\rho,
\label{adel6}
\ee
which is the fundamental equation of hydrostatic equilibrium. In the presence of
a central body, it has 
to be solved with the boundary condition
\be
\frac{dP}{dr}(R_{*})=-\rho(R_{*})\frac{GM_{*}}{R_{*}^{d-1}}.
\ee
For the isothermal equation of state 
\be
P=\rho\frac{k_B T}{m},
\label{adel2}
\ee
the foregoing equation becomes
\be
\frac{1}{r^{d-1}}\frac{d}{dr}\left ({r^{d-1}}\frac{d\ln\rho}{dr}\right )=-S_{d}\beta G m \rho
\label{adel7}
\ee
with
\be
\left (\frac{d\ln\rho}{dr}\right )(R_*)=-\beta m\frac{GM_*}{R_*^{d-1}}.
\ee
Writing $\rho(r)=\rho_{0}e^{-\psi(\xi)}$
with the variables 
$\psi$ and $\xi$ defined previously, we recover the Emden equation (\ref{emden}) and the
boundary condition from Eq. (\ref{emdenbc}).

The two descriptions are of course equivalent since the Maxwell-Boltzmann
distribution (\ref{adel}) implies the isothermal equation of state (\ref{adel2}). Indeed, defining
the density and the pressure by $\rho=\int f\, d{\bf v}$ and
$P=\frac{1}{d}\int f v^2\, d{\bf v}$, we immediately obtain Eq. (\ref{adel2}) from
Eq. (\ref{adel}). On the other hand, the condition of hydrostatic
equilibrium (\ref{adel3}) can be recovered from the Maxwell-Boltzmann distribution (\ref{adel}) as follows.
Taking the logarithmic derivative of Eq.
(\ref{dion}) we obtain
\begin{eqnarray}
\frac{\nabla\rho}{\rho}=-\beta m\nabla\Phi_{\rm tot}.
\label{adel4}
\end{eqnarray}
Using the isothermal equation of state (\ref{adel2}), we see that Eq. (\ref{adel4}) is
equivalent to Eq. (\ref{adel3}). More generally, the condition of
hydrostatic equilibrium (\ref{adel3}) is satisfied by any self-gravitating system
which is described by a distribution function that only depends on the
individual energy of the particles: $f({\bf
r},{\bf v})=f(\epsilon)$ where $\epsilon=v^2/2+\Phi({\bf r})$ (see Appendix C
of \cite{css}).

{\it Remark:} Eqs. (\ref{adel6}) and (\ref{adel7}) can
be written in
vectorial form
as
\be
\nabla\cdot \left (\frac{\nabla P}{\rho}\right )=-S_d G\rho
\ee
and
\be
\Delta\ln\rho+\frac{S_{d} G m}{k_B T}  \rho=0.
\ee
These equations are valid even if the system is not spherically symmetric.

\subsection{Equation for the mass profile}

For the isothermal equation of state (\ref{adel2}), the condition of hydrostatic equilibrium (\ref{vio}) becomes
\be
\label{emp1}
\frac{k_B T}{m}\frac{d\rho}{dr}=-\rho\frac{G(M_{*}+M(r))}{r^{d-1}}.
\ee
Substituting Eq. (\ref{eq.rhom'}) into Eq. (\ref{emp1}), we obtain an equation for the mass profile $M(r)$ of the form
\be
\label{emp3}
\frac{k_B T}{m}\frac{d}{dr}\left (\frac{M'}{S_{d}r^{d-1}}\right )=-\frac{M'}{S_{d}r^{d-1}}\frac{G(M_{*}+M(r))}{r^{d-1}}.
\ee
After simplification, we get
\be
\label{emp4}
M''-\frac{d-1}{r}M'+\beta Gm \frac{M'(M_{*}+M(r))}{r^{d-1}}=0.
\ee
This differential equation has to be solved with the boundary conditions
\be
\label{emp5}
M(R_{*})=0, \qquad M(R)=M.
\ee
The box radius $R$ is
replaced by $+\infty$ if we consider
unbounded configurations in $d=1$ and $d=2$ dimensions.

\subsection{The case of a central Dirac mass}

If there is a Dirac mass $M_*$ at $r=0$, the total gravitational potential satisfies the Poisson equation
\be
\Delta\Phi_{\rm tot}=S_d G\rho+S_d G M_* \delta({\bf r}).
\label{daine}
\ee
Indeed, $\Delta\Phi=S_d G\rho$ and $\Delta\Phi_{\rm ext}=S_d G M_* \delta({\bf r})$. This equation is valid for any ${\bf r}$. For an isothermal gas, the Boltzmann-Poisson equation becomes
\be
\label{eq.equationboltzpoiss}
\Delta\Phi_{\rm tot}=S_{d}GAe^{-\beta m \Phi_{\rm tot}}+S_d G M_* \delta({\bf
r}).
\ee
If the system is spherically symmetric, integrating the   Boltzmann-Poisson equation (\ref{eq.equationboltzpoiss}) between $0$ and $r$, we obtain
\begin{equation}
r^{d-1} \frac{d \Phi_{\rm tot}}{d r} = GM_*
+ S_d G A \int_{0}^r  e^{-\beta m \Phi_{\rm tot}(r')} \, {r'}^{d-1} \,dr'.
\end{equation}
This equation is equivalent to Newton's law (\ref{oas}). For $r\rightarrow 0$, we have
\begin{equation}
\label{dain2}
\frac{d \Phi_{\rm tot}}{d r} \sim \frac{GM_*}{r^{d-1}}\qquad (r\rightarrow 0).
\end{equation}
On the other hand, combining the condition of hydrostatic equilibrium (\ref{adel3}) with the Poisson equation (\ref{daine}) and proceeding as in Sec. \ref{sec_fy} we obtain the differential equation
\be
\nabla\cdot \left (\frac{\nabla P}{\rho}\right )=-S_d G\rho-S_d G M_*
\delta({\bf r}). 
\ee
For the isothermal equation of state (\ref{adel2}) it takes the form
\be
\Delta\ln\rho+\frac{S_{d} G m}{k_B T}  \rho=-\frac{S_d G m}{k_B T}M_*
\delta({\bf r}). 
\ee
From Eqs. (\ref{vio}), (\ref{adel2}) and (\ref{dain2}) we obtain
\be
\left (\frac{d\ln\rho}{dr}\right )(r)\sim -\beta m\frac{GM_*}{r^{d-1}}
\ee
when $r\rightarrow 0$. The other formulae established in the previous sections remain valid with $R_*=0$.

{\it Remark:} In $d=3$ dimensions,  we have $\Phi_{\rm
tot}\sim\Phi_{\rm ext}\sim -GM_*/r$ for $r\rightarrow 0$ implying that
the density $\rho\sim e^{\beta GM_* m /r}$ [see Eq. (\ref{dion})] strongly
diverges. In that case, the density profile is not
normalizable. There is no equilibrium state with the Boltzmann
distribution around a Dirac mass in $d=3$ and the gas collapses (see Appendices
\ref{sec_sgbp} and \ref{sec_dye}).

\subsection{Notations}
\label{sec_notations}

It will be convenient in the following to introduce the normalized mass and the normalized radius of the central body 
\be
\label{mu}
\mu=\frac{M_*}{M},\qquad \zeta=\frac{R_*}{R}.
\ee
Following \cite{css} we also introduce the normalized energy and the normalized  inverse temperature of the gas
\begin{eqnarray}
\label{lambdaD}
\Lambda=-\frac{ER^{d-2}}{GM^2},\qquad \eta=\frac{\beta GMm}{R^{d-2}}.
\end{eqnarray}

\section{Analytical results in one dimension}
\label{sec.analytique1d}

The density profile of a one-dimensional self-gravitating isothermal 
gas without central body in the mean field approximation (valid for $N\gg 1$) can be obtained analytically and is known for a long
time. This is the solution of the one-dimensional Boltzmann-Poisson equation. In
an infinite domain, this solution was obtained in
\cite{spitzer,camm,ledoux,pacholczyk,stodolkiewicz,rybicki,hlake,sirechavanis2002,
chavanissire2006a,chavanis2007,muller}. It is often
called the Camm \cite{camm} profile. This solution has been generalized in a
finite domain in \cite{katzlecar,sirechavanis2002,chavanis2007,muller}. On the
other hand, the thermodynamics of a one-dimensional self-gravitating gas,
including the computation of the caloric curve, has been studied by
\cite{katzlecar,sirechavanis2002} in a finite domain and by
\cite{chavanis2006,chavanissire2006a}
in an infinite domain.\footnote{Note that Rybicki \cite{rybicki} obtained the one-particle distribution function and the caloric curve $E=\frac{3}{2}(N-1)k_B T$ in an infinite domain in closed form for
any value of $N$, not only for $N\gg 1$. The exact equilibrium
statistical mechanics of one-dimensional self-gravitating systems was also 
studied by Salzberg \cite{salzberg} but he considered the usual thermodynamic
limit $N\rightarrow +\infty$ with $N/R$ fixed (in $d=1$) which is not suitable
for self-gravitating systems. The proper thermodynamic limit 
for self-gravitating systems is
discussed in Appendix A of \cite{aakin}. It corresponds to
$N\rightarrow +\infty$ with $\Lambda$, $\eta$, $\mu$ and $\zeta$
fixed. A relevant scaling is $N\rightarrow
+\infty$ with $m\sim 1/N$, $M\sim 1$, $G\sim 1$, $R\sim 1$,  $E\sim 1$ and
$T\sim 1/N$ (Gilbert scaling). 
Alternatively, one can take $N\rightarrow +\infty$ with $G\sim 1/N$, $m\sim 1$,
$M\sim N$, $R\sim 1$, $E\sim N$ and $T\sim 1$ (Kac scaling).
These scalings are valid in any dimension of space.} The virial
theorem was discussed in
\cite{sirechavanis2002,chavanis2006,chavanissire2006a,chavanissire2006b,
chavanis2007,chavanis2012,chavanisDexact,chavanismannella}. Important identities
such as the caloric curve in an infinite domain can be obtained from the virial
theorem \cite{chavanis2006,chavanissire2006a}.

These results  can be generalized so
as to account for the presence of a central body. In this section, we
solve the differential equation for the mass profile [see Eq. (\ref{emp4})]
\be
\label{eq.equationdiff1d}
M''+\beta Gm (M_{*}+M(x))M'=0
\ee
with $M(x_*)=0$ and $M(R)=M$, and in Appendix \ref{app.boltzpoisson1d} we solve
the corresponding Boltzmann-Poisson
equation. We consider successively the case of an infinite domain ($R=+\infty$) and the case of a finite domain ($R<+\infty$).
The following results are valid for an arbitrary value  of the radius $x_*$
of the central body (in one dimension we note $x_*$ for $R_*$). The case of a
central Dirac mass corresponds to $x_*=0$.

\subsection{Infinite domain}
\label{sec_uid}

In an infinite domain, it is convenient to normalize the mass profile $M(x)$ by
the
total mass
$M$ and the distance $x$ by the scale height defined by
\be
\label{eq.echelleH1d}
H=\frac{2k_{B}T}{GMm}.
\ee
Then, Eq. (\ref{eq.equationdiff1d}) can be rewritten as
\be
\label{eq.echelleH1da}
M''+2(\mu+M)M'=0,
\ee
where $M(x)$ stands for $M(x)/M$ and $x$ stands for $x/H$. We have
introduced the normalized mass of the central body from Eq. (\ref{mu}). Equation (\ref{eq.echelleH1da}) can be integrated once to yield
\be
\label{eq.equationdiffminf}
M'+2\mu M+M^{2}=\chi,
\ee
where $\chi$ is a constant. Since $M=0$ and $M'=2\rho_0 H/M$ [see Eq.
(\ref{eq.rhom'})] at
$x=x_*$, we find that
\be
\chi=\frac{2\rho_0 H}{M},
\ee
where $\rho_0=\rho(R_*)$ is the central density. On the other hand, owing to the fact that
$M \rightarrow 1$ and
$M' \rightarrow 0$ for $x \rightarrow +\infty$, we get
\be
\chi=2\mu+1.
\label{cm}
\ee
Therefore, the density of the gas at the contact with the central body is given by
\be
\label{eq.rho01d}
\rho_{0}=\frac{M}{2H}(1+2\mu).
\ee
In addition, using Eq. (\ref{cm}), the first
order differential equation
(\ref{eq.equationdiffminf}) can be rewritten as
\be
\label{diff}
\frac{dM}{(1-M)(1+2\mu+M)}=dx.
\ee
This equation is easily integrated into
\be
\label{main}
\frac{1-M}{1+2\mu+M}=Qe^{-2(\mu+1)x},
\ee
where $Q$ is a constant. The constant $Q$ is determined by the condition
that $M=0$ at $x=x_{*}$. This yields
\be
Q=\frac{e^{2(\mu+1)x_{*}}}{1+2\mu}.
\ee
We can then rewrite Eq. (\ref{main}) as
\be
\frac{1-M}{1+2\mu+M}=\frac{1}{1+2\mu}e^{-2(\mu+1)(x-x_{*})}.
\ee
Solving this equation for $M(x)$ and returning to the original variables, we
find that
\be
\frac{M(x)}{M}=(1+2\mu)\frac{e^{2(\mu+1)(x-x_{*})/H}-1}{(1+2\mu)e^{2(\mu+1)(x-x_{*})/H}+1}.
\ee
This equation  can be rewritten as
\be
\frac{M(x)}{M}=1-\frac{2(1+\mu)}{(1+2\mu)e^{2z}+1}
\ee
or, equivalently, as
\be
\label{mpriof}
\frac{M(x)}{M}+\mu=(1+\mu)\frac{(1+2\mu)e^{z}-e^{-z}}{(1+2\mu)e^{z}+e^{-z}},
\ee
where $z=(\mu+1)(x-x_{*})/H$. If we introduce the constant
\be
\label{eq.definitionc}
C=\frac{1}{2}\ln (1+2\mu)=\tanh^{-1}\left (\frac{\mu}{1+\mu}\right ),
\ee
where we used the identity
\be
\label{id}
\tanh^{-1}(x)=\frac{1}{2}\ln\left (\frac{1+x}{1-x}\right )
\ee
to obtain the second equality, we get
\begin{eqnarray}
\frac{M(x)}{M}&=&(1+\mu)\frac{e^{z+C}-e^{-z-C}}{e^{z+C}+e^{-z-C}}-\mu\nonumber\\
&=&(1+\mu)\tanh(z+C)-\mu.
\end{eqnarray}
The final expression of the mass
profile is therefore
\be
\frac{M(x)}{M}=(1+\mu)\tanh\left\lbrack (1+\mu)(x-x_{*})/H+C\right\rbrack-\mu.
\label{dain}
\ee
The corresponding density profile $\rho(x)=M'/2$ [see Eq. (\ref{eq.rhom'})]  is given by
\be
\label{eq.rhox1d}
\rho(x)=\frac{M}{2H}(1+\mu)^{2}\frac{1}{\cosh^{2}\left\lbrack (1+\mu)(x-x_{*})/H+C\right\rbrack},
\ee
where we used
\be
\tanh'(x)=\frac{1}{\cosh^2(x)}=1-\tanh^2(x).
\label{saez}
\ee
For $x=x_*$ we recover Eq. (\ref{eq.rho01d}). Therefore, Eq.
(\ref{eq.rhox1d}) can also be written as
\be
\label{eq.rhox1dnorm}
\frac{\rho(x)}{\rho_0}=\frac{(1+\mu)^{2}}{1+2\mu}\frac{1}{\cosh^{2}\left\lbrack (1+\mu)(x-x_{*})/H+C\right\rbrack}.
\ee
The case of a central Dirac mass $M_*$  is simply obtained by taking
$x_{*}=0$ in the foregoing equations. We note that the central density
$\rho_0=\rho(0)$ remains finite in that case. On
the other hand, in the absence of a central body ($M_*=0$ and $x_*=0$), we
recover
the Camm solution \cite{chavanis2007}:
\be
M(x)=M\tanh(x/H),
\ee
\be
\rho(x)=\frac{\rho_{0}}{\cosh^{2}(x/H)},
\ee
with
\be
\rho_{0}=\frac{M}{2H}.
\ee

\begin{figure}
\begin{center}
\includegraphics[width=8cm,angle=0,clip]{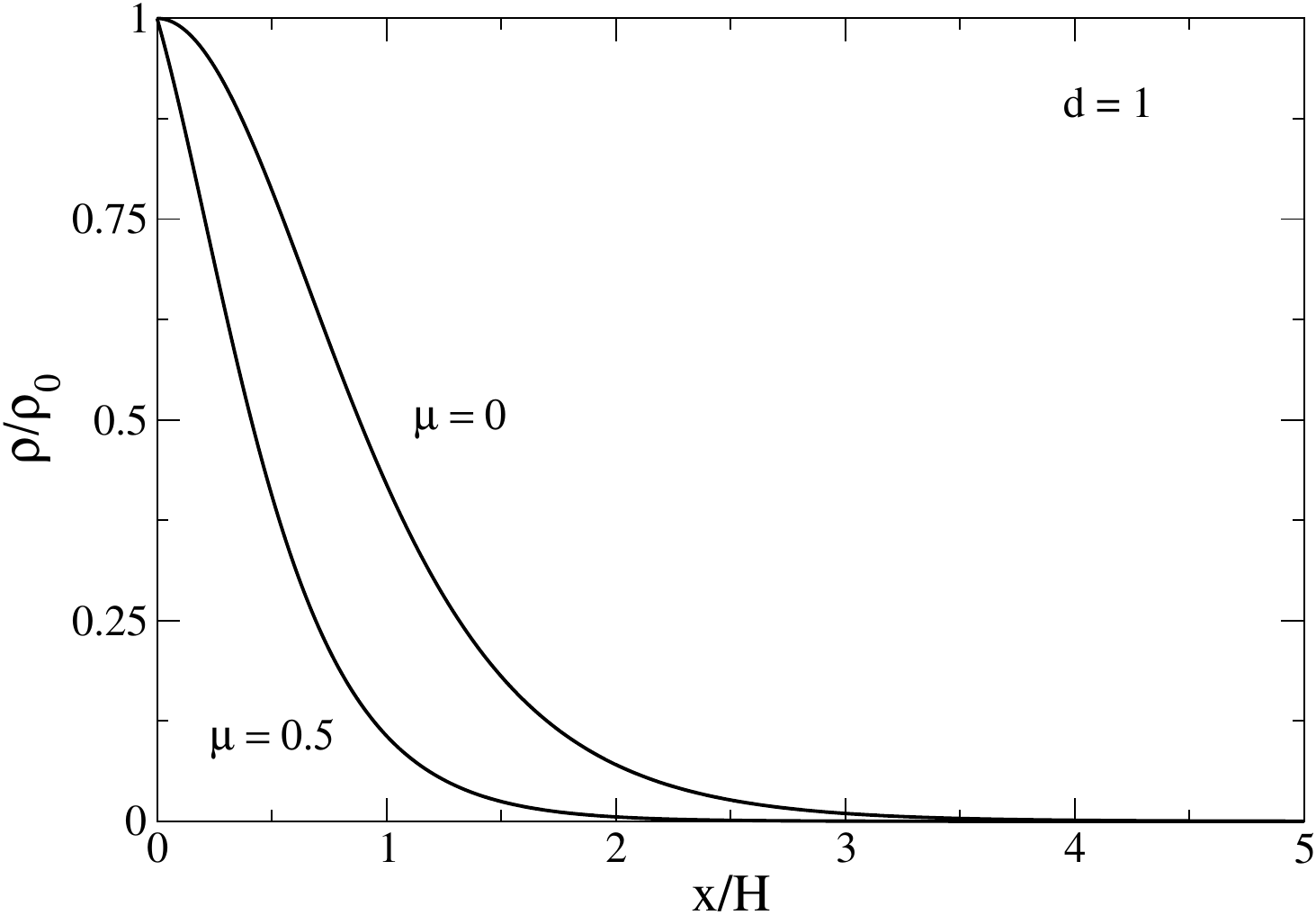}
\end{center}
\caption{\label{xrd1i} Normalized density profile of an isothermal
self-gravitating gas
surrounding a central body in an infinite domain in $d=1$ (we have taken
$\mu=0.5$ and $x_*=0$ for illustration). We have also plotted the Camm
profile corresponding to $\mu=x_*=0$. The distance has been normalized by
$H$ and
the density by $\rho_0$.}
\end{figure}

Coming back to the general solution (\ref{eq.rhox1dnorm}), we can easily obtain
its asymptotic behaviors. For $x\rightarrow +\infty$, we find that
\be
\label{asyw}
\frac{\rho(x)}{\rho_0}\sim \left\lbrack \frac{2(1+\mu)}{1+2\mu}\right\rbrack^2 e^{-2(1+\mu)x/H}.
\ee
The density decreases exponentially rapidly on a typical length scale
$H/[2(1+\mu)]$. For $x\rightarrow x_*$, we find that 
\be
\frac{\rho(x)}{\rho_0}\simeq 1-\frac{2\mu}{H}(x-x_*)+...
\ee
This is in agreement with Eq. (82) of \cite{css}.
When $\mu\neq 0$, we have a
linear increase of the density for
$x\rightarrow x_*$ as we approach the central body, leading to a central cusp ($\rho'(x_*)\neq 0$) of
typical size $H/2\mu$.
This differs from the Camm
solution (without central body) for which
\be
\frac{\rho(x)}{\rho_0}=1-\frac{x^2}{H^2}+...
\ee
In that case, the density profile is flat at the center ($\rho'(0)=0$).

We note that the normalized density  $\rho(x)/\rho_0$ as a function of
the normalized distance $X=(x-x_*)/H$  depends only on the normalized
central mass $\mu$. Some density profiles are plotted in Fig. \ref{xrd1i}.

\subsection{Finite domain}
\label{sec_elen}

In a finite domain, it is convenient to normalize the mass profile $M(x)$ by the
total
mass
$M$ and the distance $x$ by the box radius $R$. Then,
Eq. (\ref{eq.equationdiff1d}) can be rewritten as
\be
\label{dem}
M''+\eta (\mu+M)M'=0,
\ee
where $M(x)$ stands for $M(x)/M$ and $x$ stands for $x/R$. We have
introduced the normalized inverse temperature in $d=1$ dimension (see Sec. \ref{sec_notations}):
\be
\label{etau}
\eta=\beta GMmR.
\ee
Equation (\ref{dem}) can be integrated once into
\be
\label{eq.equationdiffmfini}
M'+\eta\mu M+\frac{1}{2}\eta M^{2}=\chi,
\ee
where $\chi$ is a constant. Since $M=0$ and
$M'=2\rho_{0}R/M$ [see Eq. (\ref{eq.rhom'})] at $x=x_*$, we find that
\be
\label{chiw}
\chi=2\rho_{0}\frac{R}{M}.
\ee
Therefore, $\chi$ is a measure of the central
density $\rho_0=\rho(R_*)$. This is an unknown that we will have to relate to
$\eta$, $\mu$ and $\zeta$.  The first
order differential equation (\ref{eq.equationdiffmfini}) can be rewritten as
\be
\label{fqs}
\frac{dM}{(M_{1}-M)(M+K)}=\frac{\eta}{2}dx,
\ee
where
\be
\label{mun}
M_{1}=-\mu+ \sqrt{\mu^{2}+\frac{2\chi}{\eta}}
\ee
and
\be
\label{nov}
K=\mu+ \sqrt{\mu^{2}+\frac{2\chi}{\eta}}.
\ee
We have the identities
\be
K+M_1=2\sqrt{\mu^{2}+\frac{2\chi}{\eta}},
\ee
\be
\label{peu}
K-M_1=2\mu,\qquad KM_1=\frac{2\chi}{\eta}.
\ee
Since $M(x)$ is a monotonically increasing function, Eq. (\ref{fqs}) requires
that $M_1-M(x)\ge 0$ for all $x$. Since  $M=1$ at $x=1$ this implies
that $M_1\ge 1$. Therefore, according to Eq. (\ref{mun}), we must have
\be
\chi\ge \frac{1}{2}(2\mu+1)\eta.
\ee
Taking these constraints into account, Eq. (\ref{fqs}) is readily
integrated into
\be
\label{mer}
\frac{M_{1}-M}{K+M}=Qe^{-\frac{\eta}{2}(M_{1}+K)x},
\ee
where $Q$ is a positive constant. It is determined by the
condition
that $M=0$ at $x=x_{*}$. This yields
\be
Q=\frac{M_{1}}{K}e^{\frac{\eta}{2}(M_{1}+K)x_{*}}.
\ee
We can then rewrite Eq. (\ref{mer}) as
\be
\frac{M_{1}-M}{K+M}=\frac{M_{1}}{K}e^{-\frac{\eta}{2}(M_{1}+K)(x-x_{*})}.
\ee
Solving this equation for $M(x)$ and returning to the original variables, we
find that
\be
\frac{M(x)}{M}=K\frac{e^{\frac{\eta}{2}(M_{1}+K)(x-x_{*})/R}-1}{\frac{K}{M_{1}}
e^{\frac{\eta}{2}(M_{1}+K)(x-x_{*})/R}+1}.
\ee
This equation can be
rewritten as
\be
\frac{M(x)}{M}=M_{1}\left (1-\frac{1+\frac{K}{M_{1}}}{\frac{K}{M_{1}}e^{2z}+1}\right )
\ee
or, equivalently, as
\be
\label{bou}
\frac{M(x)}{M}+\frac{1}{2}(K-M_{1})=\frac{1}{2}(K+M_1)\frac{\frac{K}{
M_{1}}e^{z}-e^{-z}}{\frac{K}{M_{1}}e^{z}+e^{-z}},
\ee
where $z=\frac{\eta}{4}(M_{1}+K)(x-x_{*})/R$. If we introduce
the constant [see Eq. (\ref{id})]
\be
\label{drc}
C=\frac{1}{2}\ln \left (\frac{K}{M_{1}}\right )=\tanh^{-1} \left
(\frac{K-M_{1}}{K+M_{1}}\right ),
\ee
we get
\begin{eqnarray}
\frac{M(x)}{M}+\frac{1}{2}(K-M_{1})&=&\frac{1}{2}(K+M_1)\frac{e^{z+C}-e^{-z-C
} } { e^{z+C}+e^{-z-C}}\nonumber\\
&=&\frac{1}{2}(K+M_1)\tanh(z+C).\qquad 
\end{eqnarray}
Recalling Eq. (\ref{peu}),  we obtain
the final expression of the mass profile
\begin{eqnarray}
\label{eq.solutionanalytique1d}
\frac{M(x)}{M}= \frac{K+M_1}{2}\tanh\left\lbrack \frac{\eta}{4}(M_{1}+K)\frac{x-x_{*}}{R}+C\right\rbrack-\mu.\nonumber\\
\end{eqnarray}
Owing to the fact that $M(R)=M$, we find that the normalized central
density $\chi$  is determined as a function of $\eta$, $\mu$ and $\zeta$ by the
complicated algebraic equation
\begin{eqnarray}
\label{alg}
1+\mu=\frac{1}{2}(K+M_1) \tanh\left\lbrack \frac{\eta}{4}(M_{1}+K)(1-\zeta)+C\right\rbrack ,\nonumber\\
\end{eqnarray}
where $M_{1}$, $K$ and 
$C$ are functions of $\chi$ defined by Eqs. (\ref{mun}), (\ref{nov}) and
(\ref{drc}). The
corresponding
density profile $\rho(x)=M'/2$ [see Eq. (\ref{eq.rhom'})] is given by
\be
\label{deky}
\rho(x)=\frac{(K+M_{1})^{2}M\eta}{16R}\frac{1}{\cosh^
{2}\left\lbrack \frac{\eta}{4}(M_{1}+K)\frac{x-x_{*}}{R}+C\right\rbrack}.
\ee
It can also be written as
\be
\label{dek}
\frac{\rho(x)}{\rho_0}=\frac{(K+M_{1})^{2}}{4KM_{1}}\frac{1}{\cosh^
{2}\left\lbrack \frac{\eta}{4}(M_{1}+K)\frac{x-x_{*}}{R}+C\right\rbrack},
\ee
where $\rho_0$ is given by Eq. (\ref{chiw}) with Eq. (\ref{alg}). 
For $x\rightarrow x_*$, we find that
\be
\label{cab}
\frac{\rho(x)}{\rho_0}\simeq 1-\eta\mu\frac{x-x_*}{R},
\ee
in agreement with Eq. (82) of \cite{css}. The gas forms a
central cusp of typical size $\epsilon=R/(\eta\mu)=1/(\beta
GmM_*)$ at the contact with the central body. The case of a
central Dirac mass is simply
obtained by taking $x_{*}=0$ in the foregoing equations. On
the other hand, in the absence of a central body ($M_*=0$ and $x_*=0$), we have
$M_{1}=K=2(\rho_{0}R/\eta M)^{1/2}$ and $C=0$, and we recover the solution
\be
\frac{M(x)}{M}= K\tanh \left (\frac{\eta}{2}\frac{Kx}{R} \right ),
\ee
\be
\label{densc}
\frac{\rho(x)}{\rho_0}=\frac{1}{\cosh^{2}\left ( \frac{\eta}{2}\frac{K
x}{R}\right )},
\ee
with
\be
\label{mus}
1= K\tanh \left (\frac{\eta}{2}K \right ), \qquad
\frac{R\rho_0}{M}=\frac{1}{4}\eta K^2,
\ee
obtained in \cite{sirechavanis2002,chavanis2007}.

\begin{figure}
\begin{center}
\includegraphics[width=8cm,angle=0,clip]{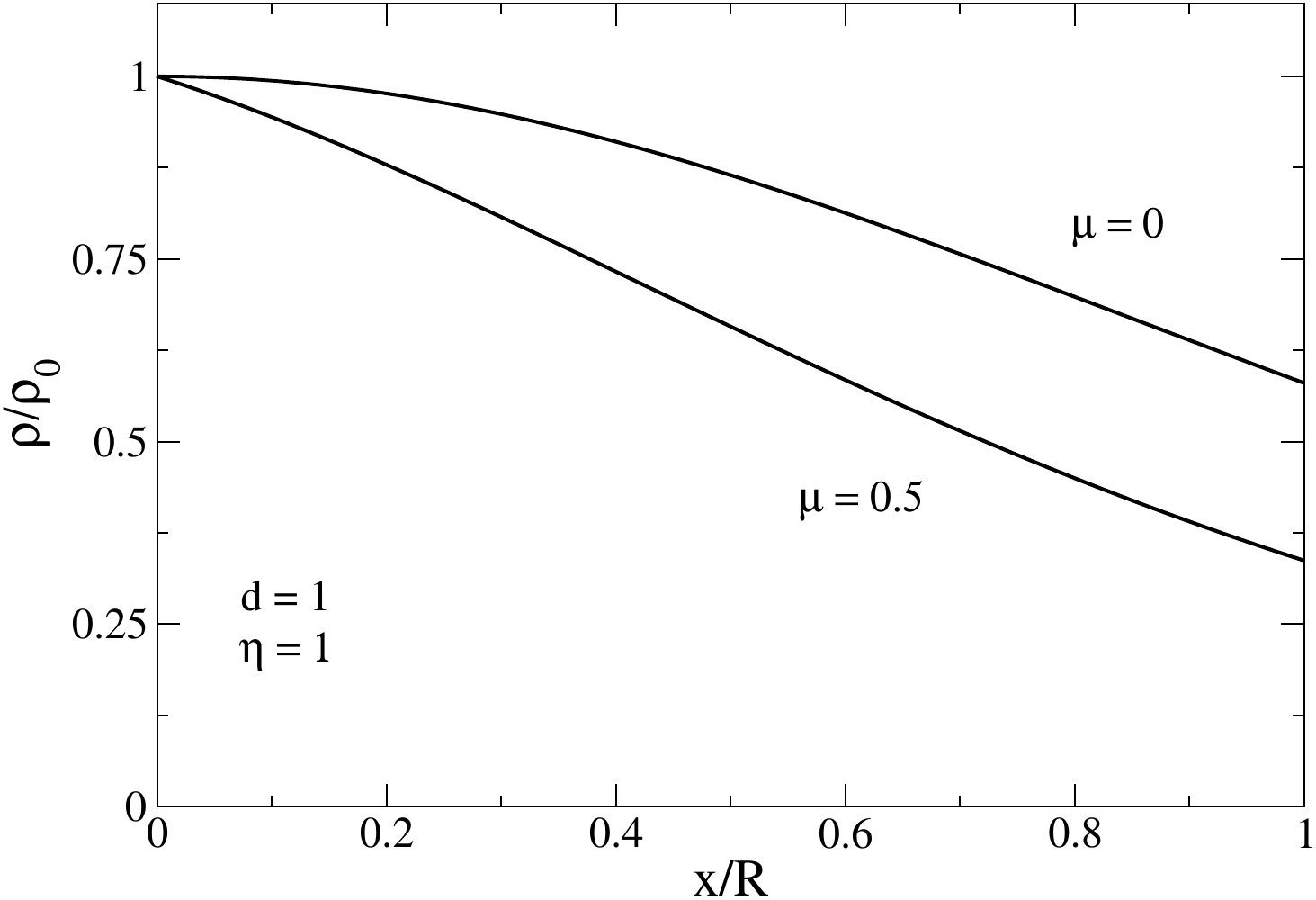}
\end{center}
\caption{\label{xrd1f} Normalized density profile of an isothermal
self-gravitating gas
surrounding a central body in a finite domain in $d=1$ (we have taken
$\eta=1$, $\mu=0.5$ and $x_*=0$ for illustration). We have also plotted the
profile corresponding to $\mu=x_*=0$. The distance has been normalized by
$R$ and
the density by $\rho_0$.}
\end{figure}

We note that the normalized density  $\rho(x)/\rho_0$ as a function of
the normalized distance $(x-x_*)/R$ depends only on the normalized inverse
temperature $\eta$, the normalized central mass $\mu$  and the
normalized radius of the central body $\zeta$. Some density profiles are plotted in Fig.
\ref{xrd1f}.

\subsection{The virial theorem}

For an isothermal equation of state $P=\rho k_{B}T/m$, the virial theorem  in
$d=1$ (see Appendix B of \cite{css})
can be written as 
\be
\label{vir}
2E_{\rm kin}-W_{\rm tot}=P(R)V-P(x_{*})V_{*},
\ee
where 
\be
\label{nv}
E_{\rm kin}=\frac{1}{2}Nk_{B}T
\ee
is the kinetic energy and $W_{\rm tot}$ is the total
potential
energy. We have also introduced the one-dimensional ``volumes'' (lengths) $V=2R$ and $V_*=2x_*$. According to Eqs.
(\ref{vir}) and (\ref{nv}), the total potential energy reads
\be
\label{wto}
W_{\rm tot}=Nk_B T+(\rho_0 x_*-\rho(R)R) \frac{2k_B T}{m}.
\ee
The total energy is then given by
\be
\label{tot}
E=E_{\rm kin}+W_{\rm tot}
\ee
with Eqs. (\ref{nv}) and (\ref{wto}). 

In an infinite domain, using the expression of $\rho_0$ from Eq.
(\ref{eq.rho01d}) and the fact that
$\lim_{R\rightarrow +\infty} R P(R)=0$
according to Eq. (\ref{asyw}), we
obtain
\be
\label{tell}
W_{\rm tot}=N k_{B}T+\frac{1}{2}(1+2\mu)GM^{2}x_{*}
\ee
and
\be
\label{top}
E=\frac{3}{2}N k_{B}T+\frac{1}{2}(1+2\mu)GM^{2}x_{*}.
\ee
Equation (\ref{top}) provides an explicit analytical expression of the caloric
curve $E(T)$. This is just an affine law. In the absence of a central body,
it reduces to \cite{chavanis2006,chavanissire2006a}
\be
E=\frac{3}{2}N k_{B}T.
\ee
This result is valid for any value of $N$ \cite{chavanis2006}.

In a finite domain, using Eqs. (\ref{chiw}) and (\ref{mu}),
we can write the total potential energy from Eq. (\ref{wto}) as
\be
\label{cul}
W_{\rm tot}=Nk_B T\left\lbrack 1+\chi(\zeta-\lambda)\right\rbrack,
\ee
where 
\be
\lambda=\frac{1}{\cal R}=\frac{\rho(R)}{\rho_0}
\ee
is the inverse density contrast. Applying Eq. (\ref{dek}) at $x=R$, we find that
\be
\label{lam}
\lambda=\frac{(K+M_{1})^{2}}{4KM_{1}}\frac{1}{\cosh^{2}\left\lbrack \frac{\eta}{4}(M_{1}+K)(1-\zeta)+C\right\rbrack}.
\ee
For given $\mu$ and $\zeta$, the
parameters $\chi$ and $\lambda$ are determined by $\eta$ according to Eqs.
(\ref{alg}) and (\ref{lam}) respectively. The caloric curve is then given by
Eqs.
(\ref{nv}), (\ref{tot}) and (\ref{cul}). In the
absence of a central body we recover the results of
\cite{sirechavanis2002}. The caloric curve is given by
\be
\label{birdy}
E=\frac{3}{2}Nk_B T-\frac{1}{2}GM^2R\frac{1}{\sinh^2\left (\frac{\eta}{2}K\right
)},
\ee
where $K$ is related to $\eta$ by Eq. (\ref{mus}). Eq. (\ref{birdy}) is
equivalent to Eq. (43) of \cite{sirechavanis2002}.

The caloric 
curve is monotonic. There is an equilibrium state for all accessible energies
$E\ge E_{\rm min}$ (with $E_{\rm min}>0$ in the presence of a central body and
$E_{\rm min}=0$ in the absence of a central body) in the microcanonical ensemble
and for all temperatures $T\ge 0$ in the canonical ensemble \cite{css}. Since
there is no turning point in the caloric curve 
all the equilibrium states are stable according to the
Poincar\'e criterion \cite{poincare,katzpoincare1,ijmpb}.\footnote{A summary of the Poincar\'e
turning point criterion for linear series of equilibria is given in Appendix C
of \cite{acepjb}.} The microcanonical and canonical ensembles are equivalent.
The caloric curve is plotted in Fig. 28 of \cite{css}. It is similar to the one
plotted in Fig. 15 of \cite{ptd} for self-gravitating fermions in a box.

{\it Remark:} The potential energy can also be obtained from the
expression (see Eq. (B31) of \cite{css})
\be
W_{\rm tot}=2G\int_{x_*}^{R} \rho(x)(M_*+M(x))x\, dx
\ee
with Eqs. (\ref{dain}) and (\ref{eq.rhox1d}) in an infinite domain ($R=+\infty$)
and with Eqs. (\ref{eq.solutionanalytique1d}) and (\ref{deky}) in a finite
domain. 
Using the identity
\be
\int \frac{\tanh(x)}{\cosh^2(x)}(x+b)\, dx=-\frac{1}{2}\frac{x+b}{\cosh^2(x)}+\frac{1}{2}\tanh(x),
\ee
we recover Eqs. (\ref{tell}) and (\ref{cul}).


\section{Analytical results in two dimensions}
\label{sec.analytique2d}

The density profile of a two-dimensional self-gravitating isothermal
gas without central body in the mean field approximation (valid for $N\gg 1$) can be obtained analytically and is known for a long
time. This is the solution of the two-dimensional Boltzmann-Poisson
equation.\footnote{This equation also arises in the context of  two-dimensional
point vortices \cite{lp,williamson,caglioti,kiessling,houchesPH,virialonsager},
plasma physics \cite{bennett,kl2}, the chemotaxis of bacterial populations
\cite{cp,hv,bkln,chavanis2007}, and string theory \cite{gervaisneveu}. Actually,
the  two-dimensional Boltzmann-Poisson equation is also known as the Liouville
equation in differential geometry and its analytical solutions have been
discovered long ago by Liouville \cite{liouville}.} In an
infinite domain, this solution was obtained in
\cite{stodolkiewicz,ostriker,alyperez,sirechavanis2002,chavanissire2006a,
chavanis2007,chavanis2012,muller}. It is often called the Ostriker profile. This
solution has been generalized in a finite domain in
\cite{katzlyndenbell,paddy,padmanabhan91,aly,sirechavanis2002,chavanis2007,
chavanis2012,muller}. On the other hand, the thermodynamics of a two-dimensional
self-gravitating gas, including the computation of the caloric curve, has been
studied by \cite{katzlyndenbell,paddy,aly,sirechavanis2002,chavanis2012} in a
finite domain and by
\cite{ostriker,alyperez,chavanissire2006a,chavanis2007,chavanis2012,
chavanisDexact,chavanismannella,muller} in an infinite domain. It exhibits a 
critical temperature
\cite{ostriker,salzberg,katzlyndenbell,paddy,padmanabhan91,abdallareza,aly,
alyperez,sirechavanis2002,chavanis2006,chavanissire2006a,chavanissire2006b,
chavanis2007,chavanis2012,chavanisDexact,chavanismannella,muller}.
The global equation of state in a finite domain is given in
\cite{salzberg,katzlyndenbell,paddy,aly,chavanis2006,chavanissire2006a,
chavanissire2006b,chavanis2007,chavanis2012,chavanisDexact,chavanismannella}.
The above results rely on a mean field approximation which is
valid for $N\gg 1$. The exact critical temperature and the
exact 
global equation of state are given in
\cite{salzberg,paddy,chavanis2006,chavanissire2006a,chavanissire2006b,
chavanis2012,chavanisDexact,chavanismannella}. The virial theorem is discussed
in
\cite{chandrafermi,ostriker,aly,alyperez,chavanis2006,chavanissire2006a,
chavanissire2006b,chavanis2007,chavanis2012,chavanismannella,chavanisDexact}. 
Important identities such as the the global equation of state and the critical
temperature can be obtained from the virial theorem.\footnote{In this sense, the
critical temperature of an isothermal self-gravitating gas in $d=2$ is implicit
in the work of Chandrasekhar and Fermi \cite{chandrafermi}.} 
 
These results can be generalized so
as to account for the presence of a central body. In this section, we solve
the differential equation for the mass profile [see Eq. (\ref{emp4})]
\be
\label{deb}
r M''-{M'}+\beta Gm {(M_{*}+M(r))M'}=0
\ee
with $M(R_*)=0$ and $M(R)=M$, and in Appendix
\ref{app.boltzpoisson2d} we solve the corresponding Boltzmann-Poisson
equation.
We first treat the case of a central Dirac mass ($R_*=0$) then the case of an
extended
central body ($R_*>0$). In each case, we consider successively the case of an 
infinite domain ($R=+\infty$)
and the case of a finite domain ($R<+\infty$). It is convenient to normalize
the mass profile $M(r)$ by the total mass $M$. Then, Eq. (\ref{deb}) can be
rewritten
as
\be
r M''-{M'}+\eta {(\mu+M)M'}=0,
\label{eq2f}
\ee
where $M(r)$ stands for $M(r)/M$, $\mu$ is the normalized mass of the
central body given by Eq. (\ref{mu}) and
\be
\label{dec}
\eta=\beta GMm
\ee
is the normalized inverse
temperature in $d=2$ dimensions (see Sec. \ref{sec_notations}). We note that it is independent of $R$.
Using $rM''=(rM')'-M'$, Eq. (\ref{eq2f}) can be integrated into
\be
r M'-(2-\eta \mu) M+\frac{1}{2}\eta M^{2}=\chi,
\label{difeqd}
\ee
where $\chi$ is a constant.

\subsection{The case of a Dirac mass}
\label{ssec.analytique2ddirac}

We first assume that we have a Dirac mass at the origin so that
$R_{*}=0$ and $M_{*}\ge 0$. The boundary condition at the origin for the gas  is
therefore $M(0)=0$. We shall also assume that $rM'(r)\rightarrow 0$ when
$r\rightarrow 0$ (we have checked afterwards that the obtained solution is consistent with this
assumption). Using  these boundary conditions in
Eq.
(\ref{difeqd}), we find that
$\chi=0$. The first order differential equation (\ref{difeqd}) can then be rewritten
as
\be
\label{rew}
\frac{dM}{(M_1-M)M}=\frac{1}{2}\eta\frac{dr}{r}
\ee
with
\be
M_1=\frac{2(2-\mu\eta)}{\eta}.
\label{m1v}
\ee
Since $M(r)$ is a monotonically increasing function, Eq. (\ref{rew}) requires
that $M_1-M(r)\ge
0$ for all $r$. Since $M(r)\rightarrow 1$ at large distances, this
implies that $M_1\ge 1$. Therefore, according to Eq. (\ref{m1v}), we must have 
\be
\eta\le \eta_c\equiv \frac{4}{1+2\mu}.
\label{inez}
\ee
Returning to the original variables, this inequality implies that an equilibrium
state may
exist only for
$T\ge T_c$ with\footnote{Alternatively, for a given temperature $T$, an
equilibrium state may exist only for $M\le M_c$ where $M_c=4k_BT/Gm-2M_*$ is a critical mass. The positivity of $M_c$ requires that $k_BT\ge GM_*m/2$.}
\be
\label{tc}
k_{B}T_{c}=\frac{GMm}{4}\left (1+2\frac{M_{*}}{M}\right ).
\ee
We note that the critical
temperature does not depend on the radius of
the central body and on the size of the system. In the absence 
of a central body ($\mu=0$), we recover the
critical temperature $k_BT_c=GMm/4$ (i.e. $\eta_c=4$) of a two-dimensional 
self-gravitating isothermal gas
\cite{sirechavanis2002,chavanis2006,chavanis2007}. Eq.
(\ref{rew}) is easily integrated into
\be
\frac{M_{1}-M}{M}=\frac{Q}{r^{F}},
\label{donm}
\ee
where $Q$ is a positive constant and
\be
F=\frac{\eta M_1}{2}=2-\mu\eta.
\label{fv}
\ee

\subsubsection{Infinite domain}
\label{sec_giua}

Let us first assume that the domain is infinite. Since $M(r)\rightarrow 1$ when $r\rightarrow
+\infty$, we find from Eq. (\ref{donm}) that $M_{1}=1$. This implies
\be
\label{eq.etainf2d}
\eta=\eta_{c}=\frac{4}{1+2\mu}.
\ee
Therefore, in an infinite domain, an equilibrium state exists at a unique
temperature
given by
Eq. (\ref{tc}). We expect that the
system evaporates for $T>T_{c}$ and that it collapses and forms
a
Dirac peak for $T\le T_{c}$. This suggests that the equilibrium states at
$T=T_c$
are
marginally stable like in the case without central body
\cite{chavanissire2006a,chavanis2007,chavanis2012} (see Appendix
\ref{sec_dye}).\footnote{Actually, they are fully
stable in the 
microcanonical ensemble and marginally stable in the canonical
ensemble \cite{chavanissire2006a,chavanis2007,chavanis2012}.} At the
critical temperature $T=T_{c}$, solving Eq.
(\ref{donm}) with 
\be
M_1=1  \quad {\rm and}\quad F=\frac{2}{1+2\mu},
\ee
and returning to the original variables, we
obtain a family of mass profiles 
\be
M(r)=\frac{Mr^{F}}{Q+r^{F}}
\ee
parameterized by $Q$. Using $\rho=M'/2\pi r$ [see Eq. (\ref{eq.rhom'})], the
corresponding density
profiles are
\be
\rho(r)=\frac{MQF}{2\pi}\frac{r^{F-2}}{(Q+r^{F})^{2}}.
\label{surprise}
\ee
For $\mu>0$, we find that the central density is always
divergent
at the origin. It behaves as a power-law
\be
\rho(r)\sim  r^{-(2-F)}\sim r^{-\frac{4\mu}{1+2\mu}} \qquad (r\rightarrow 0).
\ee
This is in agreement with Eq. (84) of \cite{css}.
At large distances, the density also decreases as a power-law but with a
different exponent:
\be
\rho(r)\sim  r^{-(2+F)} \sim r^{-\frac{4(1+\mu)}{1+2\mu}} \qquad (r\rightarrow +\infty).
\ee
We can check that the total  mass is always finite (the density is
normalizable).
For $Q\rightarrow 0$, the density profile tends to a Dirac peak
$M\delta({\bf
r})$ containing the whole mass.  For $\mu=0$, we recover the solution
without central body \cite{chavanissire2006a,chavanis2007}. This is the Ostriker \cite{ostriker} solution. This equilibrium
solution exists at a unique
(critical)
temperature 
\be
k_{B}T_{c}=\frac{GMm}{4},
\label{tay}
\ee
i.e. $\eta_{c}=4$, corresponding to
$F=2$. The mass profile is given by
\be
M(r)=\frac{\pi\rho_{0}r^{2}}{1+\frac{\pi\rho_{0}}{M}r^{2}},
\label{listen1}
\ee
and the corresponding density profile is
\be
\rho(r)=\frac{\rho_{0}}{\left (1+\frac{\pi\rho_{0}}{M}r^{2}\right )^{2}}.
\label{listen2}
\ee
Actually, we get a family of solutions parametrized by the central density
$\rho_{0}$. Contrary to the case $\mu>0$, there exist solutions with a finite  central density.
When $\rho_{0}\rightarrow +\infty$, the density profile
tends to a Dirac peak $M\delta({\bf r})$. Dynamical models based 
on the Smoluchowski-Poisson equation (appropriate to the canonical ensemble)
show that the system evaporates for $T>T_c$ and collapses towards a Dirac peak
for $T\le T_c$
\cite{sirechavanis2002,chavanissire2006a} (see Appendix \ref{sec_dye}).

\begin{figure}
\begin{center}
\includegraphics[width=8cm,angle=0,clip]{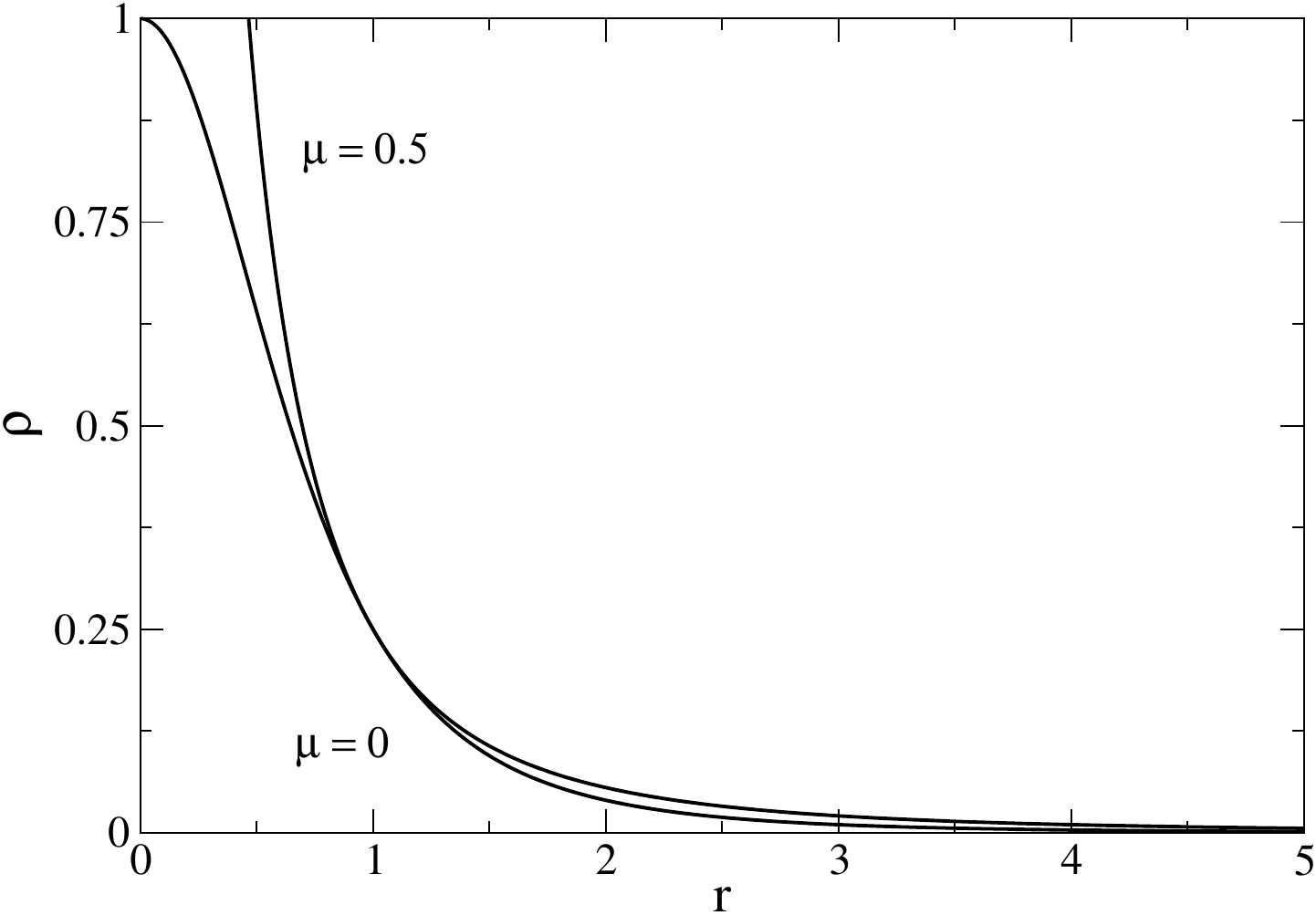}
\end{center}
\caption{\label{xrd2iD} Normalized density profile of an isothermal
self-gravitating gas
surrounding a Dirac mass in an infinite domain in $d=2$ (we have taken
$\mu=0.5$ for illustration). The distance has been normalized by $Q^{1/F}$ and
the density by $MF/(2\pi Q^{2/F})$. We have also plotted the Ostriker
profile corresponding to $\mu=0$.  The distance has been normalized by
$(M/\pi\rho_0)^{1/2}$ and
the density by $\rho_0$.}
\end{figure}

When $\mu>0$, the normalized density  $\rho(r)/[MF/(2\pi Q^{2/F})]$
as a function of the normalized distance $r/Q^{1/F}$ depends only on the
normalized
central mass $\mu$. When $\mu=0$, the normalized density  $\rho(r)/\rho_0$ as
a function
of the normalized distance $r/(M/\pi\rho_0)^{1/2}$ is universal. Some density
profiles are plotted in Fig. \ref{xrd2iD}.

In the microcanonical ensemble, 
where the control parameter is the energy, all the equilibrium states are stable
and the caloric curve is simply a horizontal line $T(E)=T_c$. In the canonical
ensemble, where the control parameter is the temperature, the equilibrium states
exist only at $T=T_c$ (with different energies) and they are marginally stable
(they collapse for $T<T_c$ and evaporate for $T>T_c$). This is similar to the
case of a classical self-gravitating isothermal gas without central body in an
infinite domain \cite{chavanissire2006a,chavanis2007}.

\subsubsection{Finite domain}
\label{sec_giub}

Let us now assume that the system is confined within a finite domain of size
$R$. Since $M=1$ when $r=R$, we find that the constant $Q$ in Eq.
(\ref{donm}) is
\be
Q=R^F(M_1-1).
\ee
We then obtain
\be
\frac{M_{1}-M}{M}=(M_1-1)\left (\frac{R}{r}\right )^{F}.
\ee
Solving this equation for $M(r)$ and returning to the original variables, we
obtain the
mass profile
\be
\frac{M(r)}{M}=\frac{M_{1}(r/R)^{F}}{M_{1}-1+(r/R)^{F}}.
\ee
This solution exists for any value of $\eta\le \eta_{c}$ (see
inequality (\ref{inez})). For $\eta>\eta_{c}$ (i.e. for $T<T_{c}$), we expect
that the system
collapses and forms a Dirac peak  like in the case without central body (see
Appendix \ref{sec_dye}). For $\eta<\eta_{c}$ (i.e. for
$T>T_{c}$), the systems has the tendency to expand  but its expansion is arrested
by the box so that an equilibrium state is finally achieved. The density profile $\rho=M'/2\pi r$ [see Eq. (\ref{eq.rhom'})] is
\be
\label{eq.rhoanalytique2dfini}
\rho(r)=\frac{FM}{2\pi R^{2}}\frac{M_{1}(M_{1}-1)(r/R)^{F-2}}{\left\lbrack
M_{1}-1+(r/R)^{F}\right\rbrack^{2}}.
\ee
For $\mu>0$, we find that the central density
is always
divergent
at the origin. It behaves as a power-law
\be
\rho(r)\sim  r^{-(2-F)}\sim r^{-\eta\mu}, \qquad (r\rightarrow 0).
\ee
This is in agreement with Eq. (84) of \cite{css}. We
can check that the total mass is always finite (the density is
normalizable). For
$\eta=\eta_{c}$ (corresponding to $M_{1}=1$), the density is a Dirac peak containing the whole mass:
$\rho({\bf
r})=M\delta({\bf r})$. For $\mu=0$, we have
$M_1=4/\eta$ and $F=2$, and 
we recover the solution without central
body \cite{sirechavanis2002,chavanis2007,chavanis2012}:
\be
\label{nou1}
M(r)=\frac{4M}{4-\eta}\frac{(r/R)^{2}}{1+\frac{\eta}{4-\eta}(r/R)^{2}},
\ee
\be
\label{nou2}
\rho(r)=\frac{4M}{\pi R^{2}(4-\eta){\left\lbrack
1+\frac{\eta}{4-\eta}(r/R)^{2}\right\rbrack^{2}}}.
\ee
This solution exists for any $\eta\le \eta_{c}=4$. Contrary to the
case $\mu>0$, the central
density is finite  and given by
\be
\label{nou3}
\rho_0=\frac{4M}{\pi R^{2}(4-\eta)}.
\ee
The density
tends to a Dirac peak $M\delta({\bf r})$ at the critical temperature
$T_c$. Dynamical models based 
on the Smoluchowski-Poisson equation (appropriate to the canonical ensemble)
show that the system tends to an equilibrium state for $T>T_c$ and collapses
towards a Dirac peak
for $T\le T_c$
\cite{sirechavanis2002,chavanissire2006a} (see Appendix \ref{sec_dye}).

\begin{figure}
\begin{center}
\includegraphics[width=8cm,angle=0,clip]{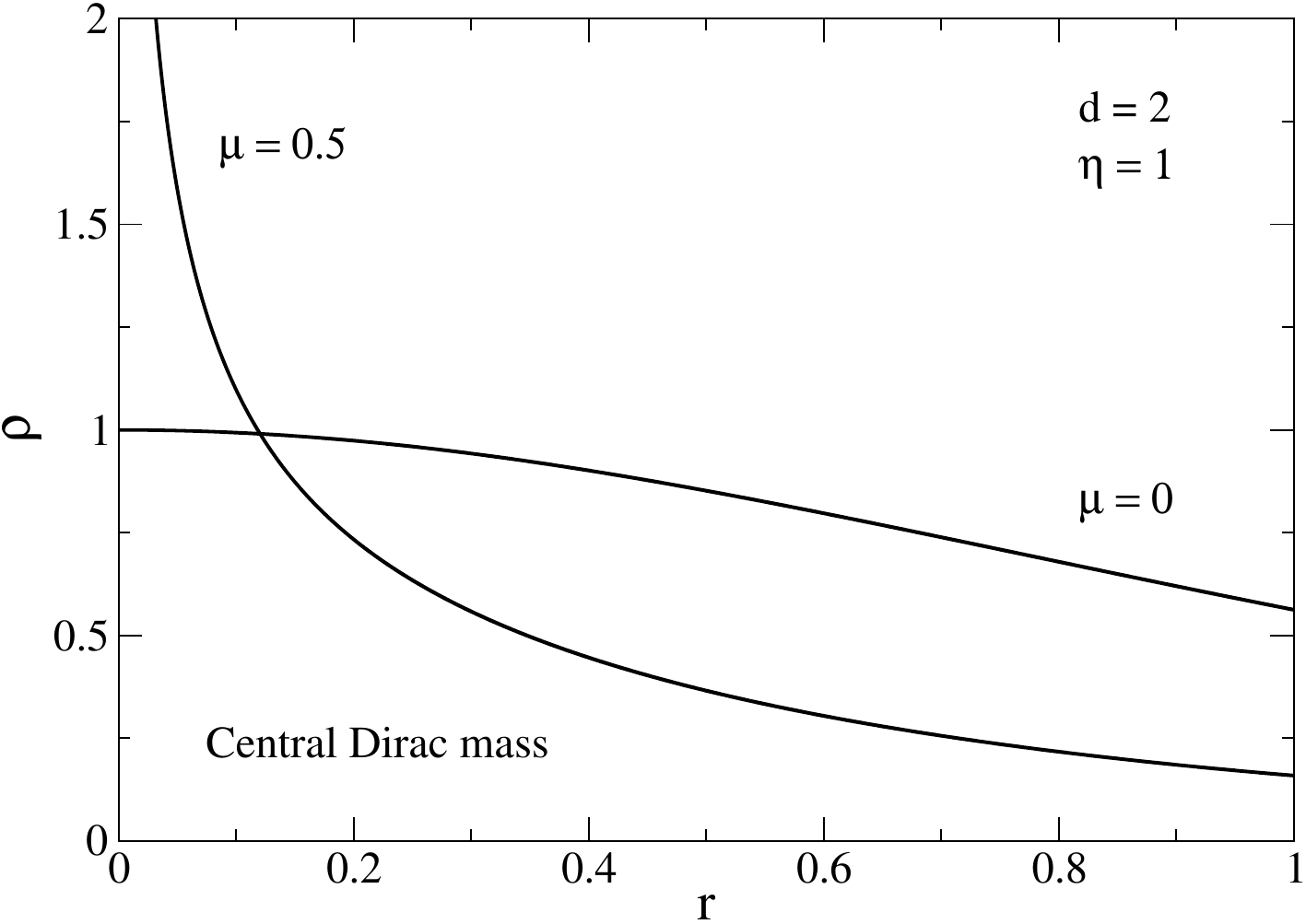}
\end{center}
\caption{\label{xrd2fD} Normalized density profile of an isothermal
self-gravitating gas
surrounding a Dirac mass in a finite domain in $d=2$ (we have taken $\eta=1$
and $\mu=0.5$ for illustration). The distance has been normalized by $R$
and
the density by $M/R^2$. We have also plotted the profile corresponding
to $\mu=0$. In that case, the density has been normalized by $\rho_0$.}
\end{figure}

When $\mu>0$, the normalized density  $\rho(r)/(M/R^2)$
as a function of the normalized distance $r/R$ depends only on the
normalized
central mass $\mu$ and the normalized temperature $\eta$. When $\mu=0$, the
normalized density $\rho(r)/\rho_0$ as
a function
of the normalized distance $r/R$ depends only on he normalized temperature
$\eta$. Some
profiles are plotted in Fig. \ref{xrd2fD}.

The caloric curve is monotonic.  In the microcanonical
ensemble, equilibrium states exist at all energies $E$. In the canonical
ensemble, equilibrium states exist only for $T\ge T_c$ (otherwise the system
collapses). Since there is no turning point in the caloric curve all the equilibrium states are stable
according to the Poincar\'e criterion . The
microcanonical and canonical ensembles are equivalent. The caloric curve is
similar in shape to the one plotted in Fig. 7 of \cite{sirechavanis2002} for a
classical self-gravitating
isothermal gas without central body in a box.

\subsection{The case of an extended central body}

We now consider the case where the central body of mass $M_*$ has a nonzero
radius
$R_{*}>0$. In that case, the boundary conditions at the origin for the gas are $M(R_{*})=0$ and
$M'(R_{*})=2\pi\rho_{0}R_{*}/M$, where $\rho_0=\rho(R_*)$ is the density of the
gas at 
the contact with the central body. As mentioned previously, we call it the central density. Using these
boundary conditions in Eq.
(\ref{difeqd}), we find that 
\be
\chi=\frac{2\pi\rho_{0}R_{*}^{2}}{M}.
\label{chh}
\ee
The parameter $\chi$ can be interpreted as a normalized central density. This is an unknown that we will have to relate to $\eta$, $\mu$ and $\zeta$. The
first order differential equation (\ref{difeqd}) can be rewritten
as
\be
\frac{dM}{(M_1-M)(M+K)}=\frac{1}{2}\eta\frac{dr}{r},
\label{gt}
\ee
where 
\be
\label{m1d}
M_{1}=\frac{2-\eta\mu+ F}{\eta},
\ee
\be
\label{kd}
K=\frac{-2+\eta\mu+ F}{\eta},
\ee
\be
\label{fd}
F=\sqrt{(2-\eta\mu)^{2}+2\eta\chi}.
\ee
We have the identities
\be
K+M_1=\frac{2F}{\eta},\qquad K-M_1=\frac{2\eta\mu-4}{\eta},
\ee
\be
\label{cime}
K M_1=\frac{2\chi}{\eta},\qquad \frac{K}{M_1}=\frac{F-2+\mu\eta}{F+2-\mu\eta}.
\ee
It is clear that $K\ge 0$. On the other hand, since $M(r)$ is a monotonically
increasing function, Eq. (\ref{gt}) requires that $M_1-M(r)\ge
0$ for all $r$. Since $M(r)\rightarrow 1$ at large distances, this
implies that $M_1\ge 1$. Therefore, according to Eqs. (\ref{m1d}) and (\ref{fd}), we
must have
\be
\chi\ge \frac{1}{2}(2\mu+1)\eta-2.
\ee
Taking these constraints into account and
integrating the differential
equation (\ref{gt}) we obtain
\be
\label{dw}
\frac{M_1-M}{M+K}=\frac{Q}{r^{F}},
\ee
where $Q$ is a positive constant.

\subsubsection{Infinite domain}
\label{sec_lum1}

In an infinite domain, using the fact that $M(r)\rightarrow 1$ when
$r\rightarrow +\infty$, we find from Eq. (\ref{dw}) that $M_{1}=1$. On the
other hand,
using the fact that $M=0$ at $r=R_{*}$, we find that
\be
Q=\frac{R_*^F}{K}.
\ee
We can then rewrite Eq. (\ref{dw}) as 
\be
\frac{1-M}{M+K}=\frac{1}{K}\left (\frac{R_*}{r}\right )^{F}.
\ee
Solving this equation for $M(r)$ and returning to the original variables, we obtain the
mass profile:
\be
\label{eq.masseinf2d}
\frac{M(r)}{M}=K\frac{(r/R_{*})^{F}-1}{1+K(r/R_{*})^{F}}.
\ee
The constants $F$ and $K$ can be determined as follows. Using Eq. (\ref{m1d}),
the condition $M_{1}=1$ yields
\be
\label{eq.Finf2d}
F=(1+\mu)\eta-2.
\ee
Then, using Eq. (\ref{fd}), we obtain
\be
\label{eq.chieta}
\chi=\frac{(2\mu+1)\eta-4}{2}.
\ee
This relation can also be
directly obtained from
Eq. (\ref{difeqd}) by using the fact that $M\rightarrow 1$ and $M'\rightarrow
0$ for $r\rightarrow +\infty$. Eq. (\ref{eq.chieta}) determines the central density
$\chi$  as a function of the inverse temperature $\eta$ (for given $\mu$). 
Then, combining  Eqs. (\ref{kd}), (\ref{eq.Finf2d}) and (\ref{eq.chieta}), we
get
\be
\label{eq.termeK2d}
K=\frac{2F}{\eta}-1=\frac{(2\mu+1)\eta-4}{\eta}=\frac{2\chi}{\eta}.
\ee
The mass profile is therefore given by Eq. (\ref{eq.masseinf2d})
where $F$ is given by (\ref{eq.Finf2d}) and $K$ by (\ref{eq.termeK2d}).
This fully determines the mass profile in terms of the inverse temperature
$\eta$ and the mass ratio $\mu$. Finally, 
the density profile $\rho(r)=M'/2\pi r$ [see Eq. (\ref{eq.rhom'})] is given by
\be
\label{comin}
\rho(r)=\frac{MKF(1+K)}{2\pi R_*^2}\frac{(r/R_{*})^{F-2}}{\left\lbrack
1+K(r/R_{*})^{F}\right\rbrack^{2}}.
\ee
It can be written as
\be
\label{eq.rhoextinf2dcc}
\frac{\rho(r)}{\rho_{0}}=\frac{(1+K)^{2}(r/R_{*})^{F-2}}{\left\lbrack
1+K(r/R_{*})^{F}\right\rbrack^{2}},
\ee
where $\rho_0$ is given by Eq. (\ref{chh}) with Eq. (\ref{eq.chieta}). 
For $r\rightarrow +\infty$, the density decreases as
\be
\label{enc1}
\frac{\rho(r)}{\rho_{0}}\sim \frac{(1+K)^{2}}{K^2}\frac{1}{(r/R_{*})^{F+2}}.
\ee
For $r\rightarrow R_*$, we have
\be
\label{enc2}
\frac{\rho(r)}{\rho_{0}}\simeq 1-\mu\eta\left (\frac{r}{R_*}-1\right ),
\ee
in agreement with Eq. (84) of \cite{css}. The gas forms a
central cusp of typical size $\epsilon=R_*/(\eta\mu)=R_*/(\beta
GmM_*)$ at the contact with the central body. For $R_*=0$ we recover the results of Sec. \ref{sec_giua}. For
$R_*=M_*=0$ we recover Eqs. (\ref{listen1}) and (\ref{listen2}).

Since
$\chi\ge 0$, an equilibrium state exists only for
\be
\eta\ge\eta_{c}\equiv  \frac{4}{2\mu+1}
\ee
i.e. $T\le T_c$. There is no equilibrium state for $\eta<\eta_{c}$
(i.e. high temperatures $T>T_{c}$) because, in an infinite domain, the
system evaporates.  For $\eta\ge \eta_{c}$ (i.e. low temperatures $T\le T_{c}$),
the system collapses but the complete collapse is arrested by the central body
so that an equilibrium state exists (by contrast, for $\mu=0$ the collapse for
$T\le T_{c}$ leads to a Dirac peak).

\begin{figure}
\begin{center}
\includegraphics[width=8cm,angle=0,clip]{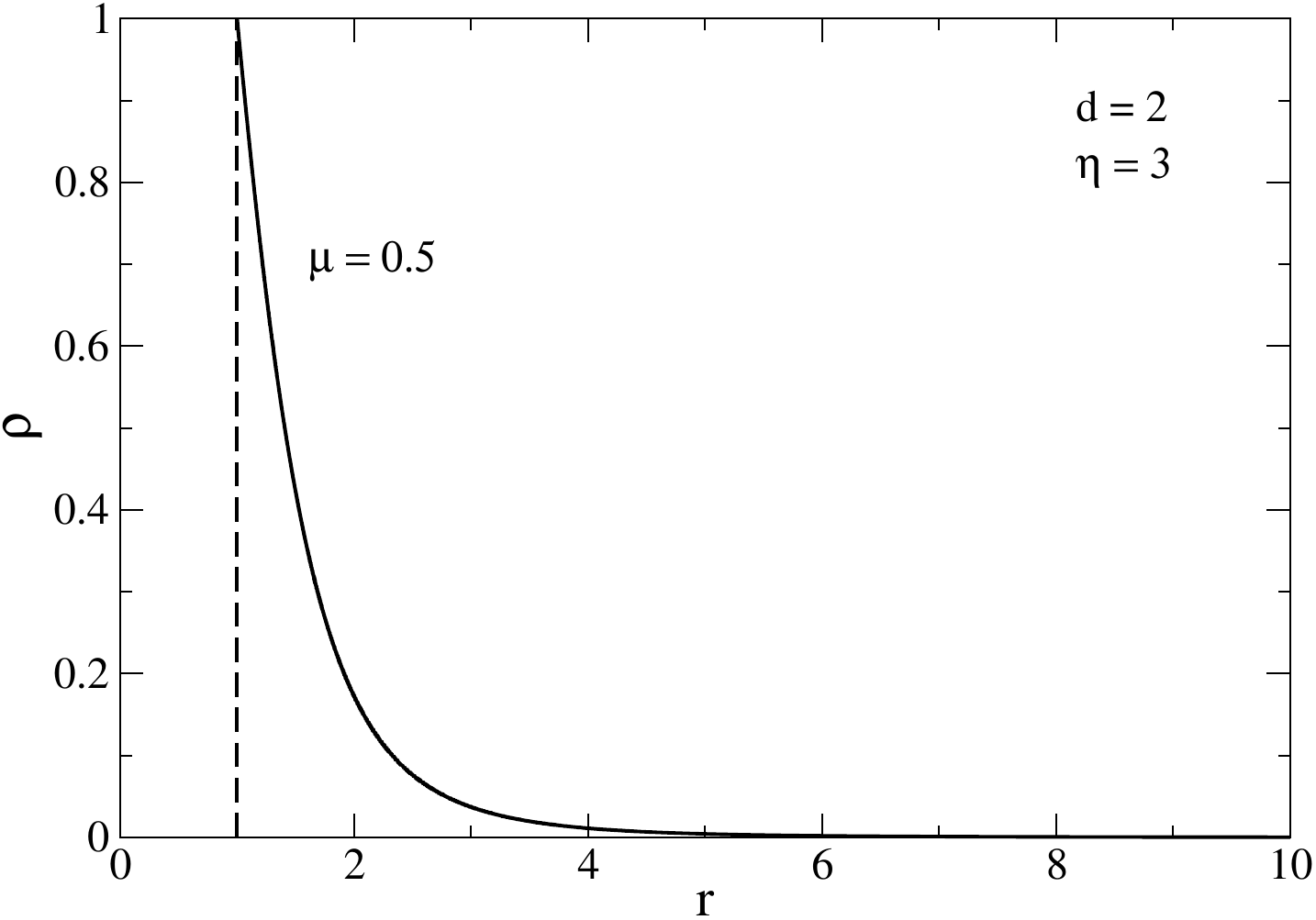}
\end{center}
\caption{\label{xrd2i} Normalized density profile of an isothermal
self-gravitating gas
surrounding a central body in an infinite domain in $d=2$ (we have taken
$\eta=3$ and $\mu=0.5$ for illustration). The distance has been normalized by
$R_*$ and
the density by $\rho_0$.}
\end{figure}

When $\mu>0$, the normalized density  $\rho(r)/\rho_0$
as a function of the normalized distance $r/R_*$ depends only on $K$ which is a function of the
normalized inverse temperature $\eta$ and normalized 
central mass $\mu$ given by Eq. (\ref{eq.termeK2d}). Some
profiles are plotted in Fig. \ref{xrd2i}.

The caloric curve is monotonic. In the microcanonical
ensemble, equilibrium states exist for all accessible energies $E\ge E_{\rm
min}$. In the canonical ensemble, equilibrium states exist only for $T\le T_c$
(for $T\ge T_c$ the system evaporates).  Since there is no turning point in the caloric curve
all the equilibrium states are stable according to the Poincar\'e criterion. The
microcanonical and canonical ensembles are equivalent.  The caloric curve  is
similar to the one plotted in Fig. 8(a) (dotted line) of \cite{kmcfermions}
 for self-gravitating fermions in an
infinite domain.

\subsubsection{Finite domain}
\label{sec_lum2}

We now consider the case where the gas is enclosed within a
box of radius $R$. Using the fact that $M=0$ at
$r=R_{*}$, we find that the constant $Q$ in Eq. (\ref{dw}) is given by
\be
Q=R_*^F\frac{M_1}{K}.
\ee
We can then rewrite Eq. (\ref{dw}) as
\be
\frac{M_1-M}{M+K}=\frac{M_1}{K}\left (\frac{R_*}{r}\right )^{F}.
\ee
Solving this equation for $M(r)$  and returning to the original variables, we
obtain the
mass profile
\be
\label{pleur}
\frac{M(r)}{M}=K\frac{(r/R_{*})^{F}-1}{1+\frac{K}{M_{1}}(r/R_{*})^{F}}.
\ee
Since $M(r)=M$ at $r=R$, we get
\be
\label{eq.m1fini2d}
M_{1}=\frac{1}{1-\frac{1+K}{K}\zeta^{F}},
\ee
where $\zeta$ is the dimensionless radius of the central body defined in Eq.
(\ref{mu}). Substituting
$M_{1}$ and $K$ from Eqs.
(\ref{m1d}) and (\ref{kd}) into Eq. (\ref{eq.m1fini2d}), we obtain a
complicated algebraic equation of the form 
\be
\label{jaz}
\frac{2-\eta\mu+F}{\eta}=\frac{1}{1-\frac{\eta(1+\mu)-2+F}{\eta\mu-2+F}\zeta^{F}}
\ee
determining $F$ as a function of $\eta$, $\mu$ and $\zeta$. 
The
constants $M_{1}$ and $K$ and the normalized central density $\chi$
are then given by Eqs. (\ref{m1d}), (\ref{kd}) and (\ref{fd}). Finally, the
density
profile $\rho(r)=M'(r)/2\pi r$ [see Eq. (\ref{eq.rhom'})] is
given by
\be
\label{bruitb}
\rho(r)=\frac{MKF}{2\pi R_*^2}\left (1+\frac{K}{M_{1}}\right )
\frac{(r/R_{*})^{F-2}}{\left\lbrack
1+\frac{K}{M_{1}}(r/R_{*})^{F}\right\rbrack^{2}}.
\ee
It can be written as
\be
\label{bruit}
\rho(r)=\rho_{0}\left (1+\frac{K}{M_{1}}\right
)^{2} \frac{(r/R_{*})^{F-2}}{\left\lbrack
1+\frac{K}{M_{1}}(r/R_{*})^{F}\right\rbrack^{2}},
\ee
where the central density $\rho_0$ is given by Eq. (\ref{chh}). For $r\rightarrow R_*$, we have
\be
\label{enc3}
\frac{\rho(r)}{\rho_{0}}\simeq 1-\mu\eta\left (\frac{r}{R_*}-1\right ),
\ee
in agreement with Eq. (84) of \cite{css}. The gas forms a
central cusp of typical size $\epsilon=R_*/(\eta\mu)=R_*/(\beta
GmM_*)$ at the contact with the central body. For $\zeta=0$ we recover the results of Sec.
\ref{sec_giub}. For
$\zeta=\mu=0$ we recover Eqs. (\ref{nou1})-(\ref{nou3}).

\begin{figure}
\begin{center}
\includegraphics[width=8cm,angle=0,clip]{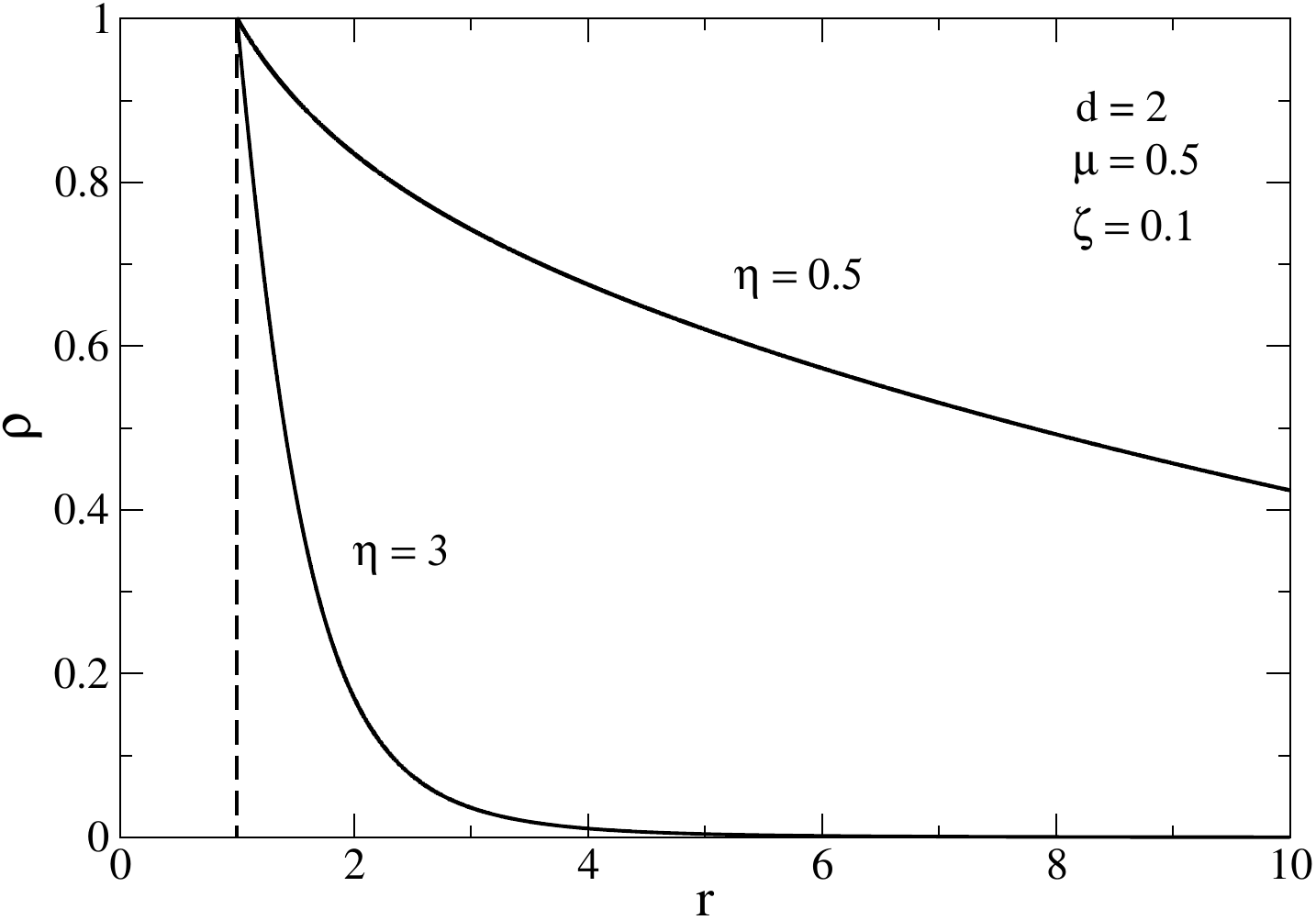}
\end{center}
\caption{\label{xrd2f} Normalized density profile of an isothermal
self-gravitating gas
surrounding a central body in a finite domain in $d=2$ (we have taken $\mu=0.5$,
$\zeta=0.1$ and $\eta=0.5, 3$ for illustration). The distance has
been normalized by
$R_*$ and
the density by $\rho_0$.}
\end{figure}

When $\mu>0$, the normalized density  $\rho(r)/\rho_0$
as a function of the normalized distance $r/R_*$ depends only on the
normalized inverse temperature $\eta$, the normalized 
central mass $\mu$ and the normalized radius of the central body $\zeta$. Some
profiles are plotted in Fig. \ref{xrd2f}.

The caloric curve is monotonic. In the microcanonical
ensemble, equilibrium states exist at all accessible energies $E\ge E_{\rm
min}$. In the canonical ensemble, equilibrium states exist at all temperatures
$T\ge 0$. Since there is no turning point in the caloric curve all the equilibrium states are stable 
according to the Poincar\'e criterion. The
microcanonical and canonical ensembles are equivalent. The caloric curve is
plotted in Fig. 27 of \cite{css}. It is similar to the one plotted in Fig. 13
of \cite{ptd} or in Fig. 8(a)
(solid lines) of \cite{kmcfermions} for self-gravitating fermions in a
box.

\subsection{The virial theorem}

Some of the preceding results can be obtained in a straightforward
manner, without having to determine the equilibrium distribution explicitly, by
using the virial theorem in two dimensions (see Appendix B of \cite{css}). For an
isothermal equation of state $P=\rho k_{B}T/m$, it takes the form
\be
E_{\rm kin}-\frac{GM^{2}}{4}\left (1+2\frac{M_*}{M}\right ) = P(R)V-P(R_*)V_*,
\label{gf1}
\ee
where
\be
E_{\rm kin}=Nk_BT
\label{gf2}
\ee
is the kinetic energy. We have introduced the two-dimensional  ``volumes''  (surfaces) $V=\pi
R^2$ and $V_*=\pi R_*^2$. Combining Eq. (\ref{gf1}) and (\ref{gf2}) we obtain
\be
Nk_B (T-T_c)= P(R)V-P(R_*)V_*
\ee
or, equivalently,
\be
Nk_B (T-T_c) = \left (\rho(R)\pi R^{2}- \rho(R_{*})\pi R_{*}^{2}\right )\frac{k_{B}T}{m},
\ee
where $T_c$ is the critical temperature from Eq. (\ref{tc}).

\subsubsection{The case of a central Dirac mass}

If we have a Dirac mass at the origin ($R_{*}=0$, $M_{*}>0$), the
virial theorem reduces to
\be
\label{vire}
PV=Nk_{B}(T-T_{c}).
\ee
We have used $\lim_{r\rightarrow 0}P(r)r^2=0$ on account of the
results
of Sec. \ref{ssec.analytique2ddirac} and we have introduced the pressure on the
box $P=P(R)=\rho(R)k_{B}T/m$. Eq. (\ref{vire})  can be viewed as the equation
of state
of a two-dimensional self-gravitating gas in the presence of a central Dirac mass. 
This result can also be obtained directly from the analytical
expression of the density profile (\ref{eq.rhoanalytique2dfini}). It
clearly shows that equilibrium solutions exist only for $T \ge T_{c}$, in
agreement with the results of Sec. \ref{sec_giub},
otherwise the pressure would be negative. If the domain is unbounded,
then $P=0$ at infinity, and we see that an equilibrium state
exists
only for $T=T_{c}$ in agreement with the results of
Sec. \ref{sec_giua}. These results generalize those obtained without 
central body where $T_c$ is given by Eq.
(\ref{tay}) \cite{salzberg,katzlyndenbell,paddy,aly,chavanis2006,
chavanissire2006a,chavanissire2006b,chavanis2007,chavanis2012,chavanisDexact,
chavanismannella}.

\subsubsection{The case of an extended central body}

If the central body has a finite extension ($R_{*}>0$), the virial theorem can
be written as
\be
\label{eq.termeviriel2d}
PV=Nk_{B}(T-T_{c})+\frac{1}{2}N\chi k_{B}T.
\ee
Eq.
(\ref{eq.termeviriel2d}) can be viewed as the equation of state of a two-dimensional self-gravitating gas in the presence of an extended central body in a finite domain.  
This result can also be directly obtained  from the analytical
expression of the density profile (\ref{bruit}). In a
finite domain, an equilibrium state exists at any temperature. In an infinite
domain ($P=0$), Eq. (\ref{eq.termeviriel2d}) implies
\be
\chi=\frac{2(T_{c}-T)}{T}.
\ee
This returns the result of Eq. (\ref{eq.chieta}). Since $\chi\ge 0$, an
equilibrium state exists only for $T \le T_{c}$. Finally, introducing the  pressure
 exerted
by the the central body on the gas, $P_{0}=P(R_{*})=\rho_0k_{B}T/m$,
we can rewrite
Eq. (\ref{eq.termeviriel2d}) under the form
\be
P_{0}V_{*}=Nk_{B}(T_{c}-T),
\ee
which can be viewed as the equation of state of a two-dimensional
self-gravitating gas in the presence of a central body in an infinite domain. 

{\it Remark:}  Eq. (\ref{eq.termeviriel2d}) can be written as
\be
\rho(R)\frac{\pi R^2}{M}=1-\frac{\eta}{\eta_{c}}+\frac{1}{2}\chi.
\ee
Using Eqs. (\ref{chh}) and (\ref{bruit}), we obtain
\be
\frac{1}{2}\chi\left (1+\frac{K}{M_{1}}\right )^{2}
\frac{1}{\zeta^F}\frac{1}{\left\lbrack
1+\frac{K}{M_{1}}\frac{1}{\zeta^{F}}\right\rbrack^{2}}=1-\frac{\eta}{\eta_{c}}
+\frac{1}{2}\chi.
\ee
This equation provides a relation between the central density $\chi$
and the inverse temperature $\eta$. It is equivalent to Eq. (\ref{jaz}).

\section{Conclusion}

The present contribution complements the results of \cite{css} where we studied the phase transitions of a self-gravitating isothermal gas around a central body in $d$ dimensions. In $d=3$ dimensions, the density profile of the gas must be computed numerically. Furthermore, equilibrium states exist only in a finite domain (i.e., if the gas is enclosed within a box). Interestingly, in $d=1$ and $d=2$ dimensions, the density profile of the gas can be obtained analytically and equilibrium states may exist (under certain conditions in $d=2$) both in finite and infinite domains. These analytical solutions generalize the well-known Camm \cite{camm} and Ostriker \cite{ostriker} solutions obtained in the absence of a central body.

There exist beautiful analogies between self-gravitating systems, the chemotaxis
of bacterial populations (Keller-Segel model) and  two-dimensional vortices
\cite{ribot}. The analogies between self-gravitating systems and  
two-dimensional vortices are discussed in detail in
\cite{these,houchesPH,tcfd,unified} and in Appendix \ref{sec_a2d}
of the present paper. The analogies between self-gravitating systems and 
bacterial populations are discussed in Appendix A of  \cite{css} and references
therein. The analogies between bacterial populations and two-dimensional point
vortices (e.g. Jupiter's great red spot) are discussed in \cite{degrad}. The equilibrium states of these different systems are determined by
the Boltzmann-Poisson equations. In the present paper and in our companion paper
\cite{css}, we have studied the equilibrium states of a self-gravitating system
surrounding a central body. The central body may represent a black hole at the
center of a galaxy or at the center of a globular cluster. It may also represent
a rocky core at the center of a giant gaseous planet (like Jupiter or Saturn).
Because of the above mentioned analogies, our study may also find applications
for bacterial populations and  two-dimensional vortices. In the case of
bacterial populations, the central body could be a supply of ``food'' that
attracts the bacteria (chemoattractant). In the case of two-dimensional
vortices, the central body could be a central vortex (punctual or extended).

\begin{appendix}

\section{Self-gravitating Brownian particles with a central body}
\label{sec_sgbp}

In this Appendix we extend the equations describing the dynamical evolution of a  gas of self-gravitating Brownian particles by taking into account the presence of a central body. 

\subsection{Smoluchowski-Poisson equations}
\label{sec_smp}

A gas of self-gravitating particles is described, in the strong friction limit $\xi\rightarrow+\infty$,  by the 
Smoluchowski-Poisson equations \cite{crs,sirechavanis2002,post}. Taking the presence of a central body into account, they can be written as 
\be
\label{sp1}
\xi\frac{\partial\rho}{\partial t}=\nabla\cdot \left
(\nabla P+\rho\nabla \Phi_{\rm tot}\right ),
\ee
\be
\label{sp2}
\Delta\Phi=S_d G\rho,
\ee
where $P=\rho k_B T/m$ is the thermal pressure and $\Phi_{\rm tot}=\Phi+\Phi_{\rm ext}$ is the total gravitational potential including the potential $\Phi$ created by the gas and the potential $\Phi_{\rm ext}$ created by the central body.\footnote{These equations are also valid for a general barotropic equation of state $P(\rho)$ and a general external potential $\Phi_{\rm ext}$ \cite{nfp}.} The equilibrium state, when it exists, is described by the equation of hydrostatic equilibrium (\ref{adel3}) reducing to the Boltzmann-Poisson equation (\ref{bpo}) in the isothermal case (see Sec. \ref{sec_fund}). The Smoluchowski equation (\ref{sp1}) can be rewritten as
\be
\label{sp3}
\xi\frac{\partial\rho}{\partial t}=\nabla\cdot \left
(\frac{k_B T}{m}\nabla\rho+\rho\nabla \Phi+\rho\frac{GM_D(t)}{r^d}{\bf r}\right ),
\ee
where we have used the expression $-\nabla\Phi_{\rm ext}=-GM_D{\bf r}/r^d$ of the force created by the central body (see Appendix B.1. of \cite{css}). We need to consider two cases: (i)  For an extended central body ($R_*>0$)  the total mass of the gas is conserved ($\dot M=0$) and $M_D=M_*$ is just the mass of the central body. (ii) For a central  Dirac mass ($R_*=0$) the total mass of the gas in $r>0$ is not conserved (except in the situations of hydrostatic equilibrium with a cuspy or singular density profile considered in Secs. \ref{sec.analytique1d} and \ref{ssec.analytique2ddirac}). Indeed, by integrating Eq. (\ref{sp3}) and assuming a regular density profile at the center and a vanishing flux at infinity or at the box surface, we obtain
 \be
\label{sp4}
\xi\frac{dM}{dt}=-S_d\rho(0,t)GM_D(t).
\ee
In that case, $M_D(t)=M_*+M_{\rm coll}(t)$ is the sum of the mass $M_*$ of the central body  and the mass  $M_{\rm coll}(t)$ of the gas that has collapsed at $r=0$ and has created a Dirac peak on its own. Clearly $\dot M_{\rm coll}=-\dot M$ on account of the conservation of the total mass. The system of equations (\ref{sp1})-(\ref{sp4}) is closed. 

{\it Remark:} These equations also describe the post-collapse regime of a gas of self-gravitating Brownian particles in the absence of a central body ($R_*=M_*=0$) when the system forms a growing Dirac mass at $r=0$ in $d\ge 2$. They have been studied in \cite{post}. When the Dirac mass is sufficiently large, we can neglect the self-gravity of the halo. In that case, we are led back to the problem of a Brownian particle submitted to a gravitational force. This problem has been studied in $d=2$ in \cite{chavanismannella}. It is equivalent to the so-called Bessel process.

\subsection{Equation for the mass profile}

For a spherically symmetric distribution of matter, the Smoluchowski equation (\ref{sp3}) reads
\be
\label{mp1}
\xi\frac{\partial\rho}{\partial t}=\frac{1}{r^{d-1}}\frac{\partial}{\partial r}\left\lbrace r^{d-1}\left
(\frac{k_B T}{m}\frac{\partial\rho}{\partial r}+\rho\frac{\partial\Phi}{\partial t}+\rho\frac{GM_D(t)}{r^{d-1}}\right )\right\rbrace.
\ee
The mass profile of the gas is defined by
\be
\label{mp2}
M(r,t)=\int_{R_*}^r\rho(r',t)S_d {r'}^{d-1}\, dr'.
\ee
Using the relations  
\be
\label{mp3}
\frac{\partial M}{\partial r}=\rho(r,t)S_d r^{d-1},\quad \frac{\partial \Phi}{\partial r}=\frac{GM(r,t)}{r^{d-1}},
\ee
we can rewrite the Smoluchowski-Poisson  equations as a single equation for the mass profile:
\begin{eqnarray}
\label{mp4}
\xi\frac{\partial M}{\partial t}=\frac{k_B T}{m}\left (\frac{\partial^2M}{\partial r^2}-\frac{d-1}{r}\frac{\partial M}{\partial r}\right )\nonumber\\
+\frac{\partial M}{\partial r}\frac{G \lbrack M(r,t)+M_D(t)\rbrack}{r^{d-1}}-S_d\rho(0,t)GM_D(t).
\end{eqnarray}
In the case of an extended central body we have $M_D=M_*$ and the last term is absent. In $d=1$, the foregoing equation reduces to
\begin{eqnarray}
\label{mp5}
\xi\frac{\partial M}{\partial t}=\frac{k_B T}{m}\frac{\partial^2M}{\partial r^2}+\frac{\partial M}{\partial r}G\lbrack M(r,t)+M_D(t)\rbrack\nonumber\\
-2\rho(0,t)GM_D(t).
\end{eqnarray}
In the absence of a central body, this equation is isomorphic to the viscous
Burgers equation (see, e.g., Sec. 5 of \cite{chavanis2007}).

\subsection{Virial theorem}

Let us derive the scalar virial theorem for the Smoluchowski-Poisson equations
in the presence of a central body (see Refs.
\cite{chavanissire2006a,chavanissire2006b}, Appendix G of
\cite{ggpp} and Appendix B.4. of \cite{ggppBH} for some
generalizations).

Taking the time derivative of the moment of inertia
\begin{eqnarray}
\label{virial1}
I=\int \rho r^2 \, d{\bf r},
\end{eqnarray}
using the Smoluchowski equation (\ref{sp1}), integrating by parts,  and noting that the boundary terms vanish, we obtain
\begin{eqnarray}
\label{virial2}
\frac{1}{2}\xi\dot I=-\int {\bf r}\cdot \nabla P\, d{\bf r}+W_{ii}+W_{ii}^{\rm ext},
\end{eqnarray}
where $W_{ii}$ is the virial of the gravitational force and  $W_{ii}^{\rm ext}$ is the virial of the external force (see Appendix B of \cite{css}).  Integrating the first term by parts we get
\begin{eqnarray}
\label{virial3}
\frac{1}{2}\xi\dot I=d\int P\, d{\bf r}-\oint P {\bf r}\cdot  d{\bf S}+W_{ii}+W_{ii}^{\rm ext}.
\end{eqnarray}
For an isothermal equation of state $P=\rho k_B T/m$ the virial theorem becomes
\begin{eqnarray}
\label{virial4}
\frac{1}{2}\xi\dot I=dN k_B T-\oint P {\bf r}\cdot  d{\bf S}+W_{ii}+W_{ii}^{\rm ext}.
\end{eqnarray}
In $d\neq 2$, we obtain
\begin{eqnarray}
\label{virial5}
\frac{1}{2}\xi\dot I=dN k_B T-\oint P {\bf r}\cdot  d{\bf S}+(d-2)W_{\rm tot},
\end{eqnarray}
where $W_{\rm tot}=W+W_{\rm ext}$ is the total potential energy (see Appendix B of \cite{css}). In $d=2$,  we obtain
\begin{eqnarray}
\label{virial6}
\frac{1}{2}\xi\dot I=2N k_B T-\oint P {\bf r}\cdot  d{\bf S}-\frac{GM^2}{2}-GM_*M.
\end{eqnarray}
Introducing the critical temperature from Eq. (\ref{tc}) the foregoing equation can be rewritten as
\begin{eqnarray}
\label{virial9}
\frac{1}{2}\xi\dot I=2Nk_B (T-T_c)-\oint P {\bf r}\cdot  d{\bf S}.
\end{eqnarray}
Assuming  that the density is uniform on the boundaries we have
\begin{eqnarray}
\label{virial7}
\oint P {\bf r}\cdot  d{\bf S}=d\lbrack P(R)V-P(R_*)V_*\rbrack.
\end{eqnarray}
The above expressions are strictly valid for an extended central body ($R_*>0$). For a central Dirac mass ($R_*=0$), $M$ must be replaced by $M(t)$, $M_*$ must be replaced by $M_D(t)$ and the boundary term at the center vanishes. At equilibrium ($\dot I=0$), we recover the results of Appendix B of \cite{css}.

{\it Remark:} In the absence of a central body, and in an infinite domain, the boundary terms disappear and Eq. (\ref{virial9}) has a simple  analytical solution $\langle r^2\rangle=4D(T)t+\langle r^2\rangle_0$ describing a diffusive motion with a gravity modified diffusion coefficient \cite{chavanissire2006a,chavanis2006,chavanisDexact,chavanis2007,chavanis2012}
\begin{eqnarray}
\label{guit}
D(T)=\frac{k_B}{\xi m}(T-T_c).
\end{eqnarray}
This solution is valid as long as there is no Dirac mass at $r=0$ (see \cite{chavanismannella} for more details).

\section{Dynamical evolution of self-gravitating Brownian particles in the strong friction limit}
\label{sec_dye}

In this Appendix, we summarize the main results obtained in previous works concerning the dynamical evolution of self-gravitating Brownian particles  described by the Smoluchowski-Poisson equations in the strong friction limit in different dimensions of space (see also Sec. 6 of \cite{chavanis2007}). These results also apply to the chemotaxis of bacterial populations described by the Keller-Segel model (see Appendix A of \cite{css}) and to the dynamical evolution of 2D Brownian point vortices (see Appendix \ref{sec_a2d}). These results have been obtained in the absence of a central body but we expect to have similar results in the presence of a central body.

\subsection{The case $d=3$}

If the system is confined within a box there exists a critical temperature $k_B T_c=GMm/(2.52 R)$ \cite{emden,aaiso}. We have the following results:

(i) When $T\le T_c$ the system undergoes a  self-similar collapse leading to a finite time singularity \cite{crs,sirechavanis2002}. The collapse time diverges at the critical point as $t_{\rm coll}\sim (T_c-T)^{-1/2}$ \cite{estimate}. The invariant profile can be obtained analytically. The density profile decreases as $r^{-2}$ when $T>0$ and as $r^{-6/5}$ when $T=0$. The central density increases as $\rho_0\sim (t_{\rm coll}-t)^{-1}$ and the core radius decreases as $r_0\sim (t_{\rm coll}-t)^{1/2}$ when $T>0$ and as  $r_0\sim (t_{\rm coll}-t)^{5/6}$ when $T=0$. The core mass $M_0(t)\sim \rho_0(t)r_0(t)^3$ tends to zero at the collapse time as $M_0(t)\sim(t_{\rm coll}-t)^{1/2}$ when $T>0$ and as $M_0(t)\sim (t_{\rm coll}-t)^{3/2}$ when $T=0$. A Dirac peak is formed in the post-collapse regime and accretes all the mass in infinite time when $T>0$ and in a finite time when $T=0$ \cite{post,Tzero}. For short times after the collapse, the Dirac mass grows as $M_D(t)\sim(t-t_{\rm coll})^{1/2}$ when $T>0$ and as $M_D(t)\sim (t-t_{\rm coll})^{3/2}$ when $T=0$. The halo outside of the Dirac peak undergoes a self-similar expansion. The density profile decreases as $r^{-2}$ when $T>0$ and as $r^{-6/5}$ when $T=0$. The central density decreases as $\rho_0\sim (t-t_{\rm coll})^{-1}$ and the core radius increases as $r_0\sim (t-t_{\rm coll})^{1/2}$ when $T>0$ and as  $r_0\sim (t-t_{\rm coll})^{5/6}$ when $T=0$.  At $T=0$ the whole evolution of the system can be obtained analytically \cite{Tzero}.

(ii) When $T>T_c$ the system may either relax towards a metastable
equilibrium state (local but not global minimum of free energy at fixed mass) or collapse as in the case $T\le T_c$. The selection between these two behaviors depends on a complicated notion of basin of attraction \cite{crs}. 

In an infinite domain, there is no equilibrium state. The system may either
collapse (as in the box case) or evaporate (see Sec. V of \cite{chavanissire2006a}) depending on the initial conditions.

\subsection{The case $d=2$}

There is a critical temperature $k_BT_c=GMm/4$ \cite{chavanis2007}.

Let us first consider the case of a system confined within a box:

(i) When $T<T_c$, the system collapses. The density profile has a core-halo
structure \cite{sirechavanis2002}. The core undergoes a  self-similar evolution. 
Its invariant profile is similar to the equilibrium profile from Eq. (\ref{listen2}). It
forms a Dirac peak of mass $(T/T_c)M$  in a finite time $t_{\rm coll}$. The core is
surrounded by  a halo with an apparent self-similar density profile
decreasing as $r^{-\alpha(t)}$ with an effective exponent $\alpha(t)$ converging
very slowly to $\alpha=2$ when $t\rightarrow t_{\rm coll}$. The post-collapse evolution has not been fully characterized. At  $T=0$ the system undergoes a  self-similar collapse leading to a finite time singularity \cite{post,Tzero}. The density profile decreases as $r^{-1}$. The central density increases as $\rho_0\sim (t_{\rm coll}-t)^{-1}$ and the core radius decreases as $r_0\sim t_{\rm coll}-t$. A Dirac peak is formed in the post-collapse regime and accretes all the mass in a finite time. For short times after the collapse, the Dirac mass grows as $M_D(t)\sim t-t_{\rm coll}$. The halo outside of the Dirac peak undergoes a self-similar expansion. The density profile decreases as $r^{-1}$. The central density decreases as $\rho_0\sim (t-t_{\rm coll})^{-1}$ and the core radius increases as $r_0\sim t-t_{\rm coll}$.  At $T=0$ the whole evolution of the system can be obtained analytically \cite{Tzero}.

(ii) When $T=T_c$, the system collapses. The density profile containing all the mass undergoes a  self-similar evolution.  Its invariant profile is similar to the equilibrium profile from Eq. (\ref{listen2}). It
forms a Dirac peak of mass $M$  in an infinite time. The central density increases exponentially rapidly
as $\rho_0\sim e^{\sqrt{2t}}$ and the core radius decreases as $r_0\sim  e^{-\sqrt{t/2}}$ \cite{sirechavanis2002}. 

(iii) When $T>T_c$, the system tends to a stable equilibrium state \cite{sirechavanis2002}.

Let us now consider the case of a system in an infinite domain:

(i) When $T<T_c$, the system collapses. The evolution is similar to the box case but
if the density of the halo decreases as $r^{-\alpha(t)}$ with $\alpha(t)\le 2$ the total mass 
diverges (at least logarithmically) so this profile cannot be valid up to infinity.

(ii) When $T=T_c$, the system collapses. The density profile containing all the mass undergoes a  self-similar evolution.  Its invariant profile is similar to the equilibrium profile from Eq. (\ref{listen2}). It
forms a Dirac peak of mass $M$  in an infinite time. The central density increases
slowly as $\rho_0\sim \ln t$ and the core radius decreases as $r_0\sim  1/\sqrt{\ln t}$ \cite{chavanissire2006a}. The
system must also eject a 
small mass at large distances so as to satisfy the conservation of the moment of inertia $I$ [see Eq. (\ref{virial9})]
when the Dirac peak forms.  There is also a family of equilibrium states with the density profile from Eq. (\ref{listen2}) parametrized by the central density $\rho_0$.  They have the same free energy \cite{chavanis2007,chavanis2012}. They have an infinite moment of inertia $I=\infty$ (logarithmic
divergence) except when the density profile is a Dirac ($\rho_0=\infty$). These equilibrium solutions are marginally stable.

(iii) When $T>T_c$, the system evaporates in a self-similar
manner \cite{chavanissire2006a}. The particles have a ``diffusive'' motion with an effective diffusion coefficient given by Eq. (\ref{guit}).

{\it Remark:} When $T<T_c$ the post-collapse regime has been studied in Sec. 5 of \cite{chavanismannella} in the case where the Dirac peak is so massive that the self-gravity of the gas can be neglected (see the Remark at the end of Appendix \ref{sec_smp}).

\subsection{The case $d=1$}

There is a stable equilibrium state at any temperature and the system relaxes towards this equilibrium state. Its evolution can be mapped on the Burgers equation which has an explicit solution (see Sec. 5 of \cite{chavanis2007}). At  $T=0$ the system undergoes a  self-similar collapse leading to a finite time singularity \cite{post,Tzero}. The density profile decreases as $r^{-2/3}$. The central density increases as $\rho_0\sim (t_{\rm coll}-t)^{-1}$ and the core radius decreases as $r_0\sim (t_{\rm coll}-t)^{3/2}$. A Dirac peak is formed in the post-collapse regime and accretes all the mass in a finite time. For short times after the collapse, the Dirac mass grows as $M_D(t)\sim (t-t_{\rm coll})^{1/2}$. The halo outside of the Dirac undergoes a self-similar expansion. The density profile decreases as $r^{-2/3}$. The central density decreases as $\rho_0\sim (t-t_{\rm coll})^{-1}$ and the core radius increases as $r_0\sim (t-t_{\rm coll})^{3/2}$. At $T=0$ the whole evolution of the system can be obtained exactly analytically  \cite{Tzero}. This is similar to the formation of singular shocks in inviscid one-dimensional (Burgers) turbulence.

\section{Analogy between self-gravitating systems and two-dimensional point vortices}
\label{sec_a2d}

The equations governing the dynamics of an inviscid flow are the equation of continuity and the Euler equation:
\be
\frac{\partial\rho}{\partial t}+\nabla\cdot (\rho {\bf u})=0,
\label{app1}
\ee
\be
\frac{\partial {\bf u}}{\partial t}+({\bf u}\cdot \nabla){\bf u}=-\frac{1}{\rho}\nabla P.
\label{app2}
\ee
For an incompressible flow, the equation of continuity reduces to the condition $\nabla\cdot {\bf u}=0$ stating that the velocity field is divergenceless.
If, in addition, the flow is two-dimensional, this condition can be written as $\partial_x u+\partial_y v=0$, where $(u,v)$ are the components of the velocity. According to the Schwarz theorem, there exists a streamfunction $\psi$ such that 
$u=\partial_y\psi$, $v=-\partial_x\psi$ or, equivalently 
\be
{\bf u}=-{\bf z}\times \nabla\psi,
\label{app2b}
\ee
where ${\bf z}$ is a unit vector normal to the flow. The vorticity $\nabla\times {\bf u}=\omega{\bf z}$ with  $\omega=\partial_x v-\partial_y u$
is directed along the vertical axis. According to the Stokes formula, the circulation of the velocity along a closed curve (C) delimiting a domain area (S) is
\be
\Gamma=\oint_{(C)}{\bf u}\cdot d{\bf l}=\int_{(S)}\omega\, d{\bf r}.
\label{app3}
\ee
Taking the curl of Eq. (\ref{app2b}), we find that the vorticity is related to the stream function by a Poisson equation
\be
\Delta\psi=-\omega,
\label{app4}
\ee
where $\Delta=\partial^2_{xx}+\partial^2_{yy}$ is the Laplacian operator in two dimensions. In an unbounded domain, this equation can be written in integral form as
\be
\psi({\bf r},t)=-\frac{1}{2\pi}\int\omega({\bf r}',t)\ln|{\bf r}-{\bf r}'|\, d{\bf r}'
\label{app5}
\ee
and the velocity field can be expressed in terms of the vorticity as
\be
{\bf u}({\bf r},t)=\frac{1}{2\pi}{\bf z}\times \int\omega({\bf r}',t)\frac{{\bf r}-{\bf r}'}{|{\bf r}-{\bf r}'|^2}\, d{\bf r}'.
\label{app6}
\ee
In a bounded domain, Eq. (\ref{app5}) must be modified so as to take into account vortex images. The impermeability condition implies that $\psi$ is constant on the boundary and we can take $\psi=0$ by convention. Taking the curl of Eq. (\ref{app2}), the pressure term disappears due to the incompressibility condition and the Euler equation becomes
\be
\frac{\partial\omega}{\partial t}+{\bf u}\cdot\nabla\omega=0.
\label{app7}
\ee
This corresponds to the transport of the vorticity $\omega$ by the velocity field ${\bf u}$. It is easy to show that the flow conserves circulation $\Gamma$ and the energy
\be
E=\int\frac{{\bf u}^2}{2}\, d{\bf r}.
\label{app8}
\ee
Using Eqs. (\ref{app2b}) and (\ref{app4}), one has successively
\be
E=\frac{1}{2}\int (\nabla\psi)^2\, d{\bf r}=-\frac{1}{2}\int \psi\Delta\psi\, d{\bf r}=\frac{1}{2}\int \omega\psi\, d{\bf r},\label{app9}
\ee
where the second equality is obtained by a part integration with the condition $\psi=0$ on the boundary. Therefore, $E$ can be interpreted either as the kinetic energy of the flow (see Eq. (\ref{app8})) or as a potential energy of interaction between vortices (see Eq. (\ref{app9})).

We shall consider the situation in which the velocity is created by a collection of $N$ point vortices. In that case, the vorticity field can be expressed as a sum of $\delta$-functions in the form
\be
\omega({\bf r},t)=\sum_{i=1}^N \gamma_i \delta({\bf r}-{\bf r}_i(t)),
\label{app10}
\ee
where ${\bf r}_i(t)$ denotes the position of point vortex $i$ at time $t$ and
$\gamma_i$ is its circulation.\footnote{When $\omega({\bf r},t)$ is expressed as
a sum of $\delta$-functions, Eq. (\ref{app7}) can be interpreted 
as the counterpart of the Klimontovich \cite{klimontovich} equation  in
plasma physics. In the collisionless regime, the ensemble average vorticity
$\langle\omega\rangle({\bf r},t)$ satisfies an equation of the form of Eq.
(\ref{app7}) which can be interpreted as the counterpart of the Vlasov
\cite{vlasov1,vlasov2} equation in plasma physics.} According to Eqs.
(\ref{app6}) and (\ref{app10}), the velocity of a point vortex is equal to the
sum of the velocities produced by the $N-1$ other vortices, i.e.
\be
{\bf V}_i=\frac{d{\bf r}_i}{dt}=-{\bf z}\times\nabla\psi({\bf r}_i)=-\sum_{j\neq i}\frac{\gamma_j}{2\pi}{\bf z}\times \frac{{\bf r}_j-{\bf r}_i}{|{\bf r}_j-{\bf r}_i|^2}.
\label{app11}
\ee
We note that, contrary to a material particle, a point vortex has no inertia. It
is moved by the presence of the other point vortices. This corresponds to the
conception of motion according to Aristote and Descartes (velocity proportional
to ``force'') as compared to Newton (acceleration proportional to
force).

As emphasized by Kirchhoff \cite{kirchhoff}, the above dynamics can be cast in a Hamiltonian form
\be
\gamma_i\frac{dx_i}{dt}=\frac{\partial H}{\partial y_i},\qquad \gamma_i\frac{dy_i}{dt}=-\frac{\partial H}{\partial x_i},
\label{app12}
\ee
\be
H=-\frac{1}{2\pi}\sum_{i<j}\gamma_i\gamma_j \ln |{\bf r}_i-{\bf r}_j|,
\label{app13}
\ee
where the coordinates $(x,y)$ of the point vortices are canonically conjugate. These equations of motion still apply when the fluid is restrained by boundaries, in which case the Hamiltonian (\ref{app13}) is modified so as to allow for vortex images, and may be constructed in terms of Green's functions depending on the geometry of the domain. Since $H$ is not explicitly time dependent, it is a constant of the motion and it represents the potential energy of the point vortices. The other conserved quantities are the circulation $\Gamma=\sum_i\gamma_i$, the angular momentum $\sum_i\gamma_i r_i^2$, and the impulse $\sum_i\gamma_i {\bf r}_i$. Note that the Hamiltonian (\ref{app13}) does not involve a ``kinetic'' energy of the point vortices in the usual sense (i.e., a quadratic term $p_i^2/2m$). This is is related to the particular circumstance that a point vortex is not a material particle. Indeed, an isolated vortex remains at rest contrary to a material particle which has a rectilinear motion due to its inertia. Point vortices form therefore a very peculiar Hamiltonian system. Note also that the Hamiltonian of point vortices can be either positive or negative (in the case of vortices of different signs) whereas the kinetic energy of the flow is necessarily positive.

The statistical mechanics of two-dimensional point vortices was first considered by Onsager \cite{onsager}. He showed the existence of negative temperature states at which point vortices of the same sign agglomerate to form large-scale structures similar to the cyclones and anti-cyclones observed in the atmosphere. His approach was developed quantitatively by Joyce and Montgomery \cite{jm} and Pointin and Lundgren \cite{pl,lp} in a mean field approximation which becomes exact in a proper thermodynamic limit $N\rightarrow +\infty$ with $\gamma\sim 1/N$. At statistical equilibrium, the ensemble average vorticity is related to the stream function by a  Boltzmann distribution of the form\footnote{We restrict ourselves to one species of point vortices for simplicity.}
\be
\langle \omega\rangle=Ae^{-\beta\gamma\psi},
\label{app14}
\ee
where $\beta$ is the inverse temperature. Combined with the Poisson equation (\ref{app4}) one obtains the Boltzmann-Poisson equation
\be
\Delta\psi=-Ae^{-\beta\gamma\psi}.
\label{app15}
\ee
This equilibrium state can be obtained by maximizing the Boltzmann entropy
\be
S=-\int \frac{\langle \omega\rangle}{\gamma}\ln \frac{\langle \omega\rangle}{\gamma}\, d{\bf r}
\label{app16}
\ee
at fixed circulation and energy. The first variations can be written as $\delta
S-\beta\delta E-\alpha\delta \Gamma=0$, where $\alpha$ and $\beta$ are Lagrange
multipliers taking the constraints (conservation of circulation and energy) into account. They lead to the Boltzmann
distribution from Eq. (\ref{app14}). These equations are similar to those
obtained in the statistical mechanics of 
self-gravitating systems \cite{houchesPH}.

In a model of 2D Brownian point vortices \cite{bv}, the evolution of the
mean vorticity is described for $N\rightarrow +\infty$ by a Fokker-Planck
equation of the form
\be
\frac{\partial\langle \omega\rangle}{\partial t}+{\bf u}\cdot \nabla\langle \omega\rangle=\nabla \cdot \left\lbrack D\left (\nabla\langle \omega\rangle+\beta \gamma \langle \omega\rangle \nabla\psi\right )\right\rbrack,
\label{app17}
\ee
\be
-\Delta\psi=\langle \omega\rangle,
\label{app18}
\ee
which is similar to the Smoluchowski-Poisson 
equations of self-gravitating Brownian particles or to the Keller-Segel model of
bacterial chemotaxis (see Appendix A of \cite{css}). This equation relaxes
towards the Boltzmann distribution (\ref{app14}) of statistical equilibrium. On
the other hand, the kinetic theory of 2D Hamiltonian point vortices (Onsager's
model) gives at the order $1/N$ a kinetic equation of the Landau or
Lenard-Balescu form which does not relax in general towards the Boltzmann distribution. We
need to take into account three-body correlations at the order $1/N^2$ to
achieve convergence towards statistical equilibrium \cite{Kvortex2023}.

\section{Boltzmann-Poisson equation in one dimension}
\label{app.boltzpoisson1d}

In $d=1$ dimension, the Emden equation (\ref{emden}) reads
\be
\label{eq.eqdiffboltzpoiss1d}
\frac{d^{2}\psi}{d\xi^{2}}
= e^{-\psi},
\ee
where
\be
\xi=(2\beta Gm\rho_0)^{1/2}x
\ee
is the rescaled distance. In the presence of a central body, this equation has to be solved
with the boundary conditions
\bea
\label{eq.conditionlimitexi01dcc}
\psi_{0} = 0 \quad {\rm and}\quad
\psi'_{0} = \frac{\eta_0}{\xi_0}
\eea
at $\xi=\xi_0$, where $\xi_0$ and $\eta_0$ are defined by
\be
\label{xieta}
\xi_{0}=(2G\rho_{0}\beta m)^{1/2}x_{*}, \qquad \eta_0=\beta GmM_*x_*.
\ee
We note the relation
\be
\label{mars1}
\xi=\xi_0\frac{x}{x_*}.
\ee
Eq. (\ref{eq.eqdiffboltzpoiss1d}) describes the motion of a fictive particle of
unit mass in a
potential
$V(\psi)=e^{-\psi}$, where $\psi$ plays the role of the position and $\xi$
the role of the time. The particle starts at $\xi=\xi_{0}$ from the position
$\psi_{0}=0$ with the velocity $\psi'_{0}=\eta_{0}/\xi_{0}$. The energy of
the particle
\be
\label{eq.termeE1dcc}
E=\frac{1}{2}\left (\frac{d\psi}{d\xi}\right )^{2}+e^{-\psi}
\ee
is conserved. Using the initial condition (\ref{eq.conditionlimitexi01dcc}) at $\xi=\xi_{0}$, we find that
\be
\label{eq.termeExi01dcc}
E=\frac{1}{2}\left (\frac{\eta_{0}}{\xi_{0}}\right )^{2}+1.
\ee
The particle descends the potential so that $d\psi/d\xi>0$. Therefore,
the solution of Eq. (\ref{eq.eqdiffboltzpoiss1d}) is obtained by
integrating the first order differential equation
\be
\frac{d\psi}{\sqrt{2(E-e^{-\psi})}}=d\xi.
\ee
Its solution is
\be
\tanh^{-1}\sqrt{1-\frac{1}{E}e^{-\psi}}=\sqrt{\frac{E}{2}}(\xi-\xi_{0})+C,
\ee
where $C$ is a constant of integration.
Using the initial condition at
$\xi=\xi_{0}$, we find that the constant of integration is given by
[see Eq. (\ref{id})]
\be
C=\tanh^{-1}\sqrt{1-\frac{1}{E}}=\frac{1}{2}\ln\left
(\frac{1+\sqrt{1-1/E}}{1-\sqrt{1-1/E}}\right
)
\label{valc}
\ee
or, equivalently, by [see Eq. (\ref{saez})] 
\be
C=\cosh^{-1}(\sqrt{E}).
\ee
Therefore, the solution of Eq. (\ref{eq.eqdiffboltzpoiss1d}) is 
\be
\label{eq.profildensite1dcc}
e^{-\psi}=\frac{E}{\cosh^{2}\left \lbrack\sqrt{\frac{E}{2}}(\xi-\xi_{0})+C\right \rbrack}.
\ee
For $\xi\rightarrow +\infty$, we have
\be
e^{-\psi}\sim 4E e^{-\sqrt{2E}\xi}.
\ee
For $\xi\rightarrow \xi_0$, we have
\be
e^{-\psi}\simeq 1-\frac{\eta_0}{\xi_0}(\xi-\xi_0)+...,
\ee
in agreement with Eq. (40) of \cite{css}. The gas forms a
central cusp of typical size $\xi_0/\eta_0$ at the contact with the central body.
Without the central body ($\xi_0=\eta_0=0$),
we find that $E=1$ and $C=0$, 
returning the Camm
solution \cite{sirechavanis2002,chavanis2007}
\be
e^{-\psi}=\frac{1}{\cosh^{2}(\xi/\sqrt{2})}.
\ee

We now have to determine the central density $\rho_0$ as a function of $M$ and
$T$. To that purpose, we will use Newton's law which can be written in
$d=1$ as [see Eq. (\ref{mxi})]
\be
\label{eq.theogaussxi1d}
\frac{d\psi}{d\xi}=\frac{\eta_0}{\xi_0}\left\lbrack
1+\frac{M(\xi)}{M_*}\right\rbrack.
\ee
The solution then depends whether we work in an infinite domain or in a finite
domain (box).

\subsection{Infinite domain}

In an infinite domain, Eq.
(\ref{eq.theogaussxi1d}) gives the identity
\be
\label{lim}
\lim_{\xi\rightarrow +\infty}\psi'(\xi)=\frac{\eta_0}{\xi_0}\left
(1+\frac{1}{\mu}\right ).
\ee
Taking the limit $\xi\rightarrow +\infty$ in Eq.
(\ref{eq.termeE1dcc}) and using Eq. (\ref{lim})  and the fact that
$e^{-\psi}\rightarrow 0$ for $\xi\rightarrow +\infty$ according to Eq.
(\ref{eq.profildensite1dcc}), we obtain
\be
\label{cel}
E=\frac{1}{2}\frac{\eta_0^2}{\xi_0^2}\left
(1+\frac{1}{\mu}\right )^2.
\ee
From Eqs. (\ref{eq.termeExi01dcc}) and (\ref{cel}), we find that
\be
\label{eq.expressionE1d0}
E=\frac{(1+\mu)^{2}}{2\mu+1}
\ee
and
\be
\label{pow}
\frac{1}{2}\left (\frac{\eta_{0}}{\xi_{0}}\right )^{2}=\frac{\mu^2}{2\mu+1}.
\ee
On the other hand, we have [see Eq. (\ref{piano})] 
\be
\label{lov}
\frac{\eta_{0}}{\xi_{0}}=\left (\frac{\beta G m}{2\rho_0}\right )^{1/2}M_*.
\ee
From Eqs. (\ref{pow}) and (\ref{lov}) we obtain
\be
\label{eq.rho0boltzpoiss1dcc}
\rho_{0}=\frac{M}{2H}(1+2\mu),
\ee
where $H$ is defined by Eq. (\ref{eq.echelleH1d}). This returns Eq. (\ref{eq.rho01d}). Using Eqs.
(\ref{rhopsi}), (\ref{eq.profildensite1dcc}), (\ref{eq.expressionE1d0})
and (\ref{eq.rho0boltzpoiss1dcc}), the density profile is given by
\be
\label{bb}
\frac{\rho(x)}{\rho_{0}}=\frac{(1+\mu)^{2}}{2\mu+1}\frac{1}{\cosh^{2}
\left\lbrack
(1+\mu)(x-x_{*})/H+C\right\rbrack},
\ee
where $C$ is determined by
\be
C=\tanh^{-1}\left (\frac{\mu}{1+\mu}\right )=\frac{1}{2}\ln (1+2\mu)
\ee
or, equivalently, by
\be
C=\cosh^{-1}\left (\frac{1+\mu}{\sqrt{2\mu+1}}\right ).
\ee
This returns Eq. (\ref{eq.rhox1dnorm}) with Eq. (\ref{eq.definitionc}). Finally,
using Eqs. (\ref{mr}), (\ref{saez}) and (\ref{bb}), 
the mass profile is given by
\be
\frac{M(x)}{M}=(1+\mu)\tanh\left\lbrack
(1+\mu)(x-x_{*})/H+C\right\rbrack-\mu.
\ee
This returns Eq. (\ref{dain}). We have therefore recovered the results of Sec.
\ref{sec_uid}.

\subsection{Finite domain}

In a finite domain, Eq.
(\ref{eq.theogaussxi1d}) applied at $r=R$ gives
\be
\label{limb}
\psi'(\alpha)=\frac{\eta_0}{\xi_0}\left
(1+\frac{1}{\mu}\right ),
\ee
where
\be
\label{alpha}
\alpha=(2\beta Gm\rho_0)^{1/2}R
\ee
is the normalized box radius. From Eqs. (\ref{xieta}) and  (\ref{alpha}) we
obtain 
\be
\label{cry}
\xi_0=\alpha\zeta,
\ee
where $\zeta$ is defined in Eq. (\ref{mu}). Applying  Eq.
(\ref{eq.termeE1dcc}) at $\xi=\alpha$ and using Eq. (\ref{limb}) we get
\be
\label{mar2}
E=\frac{1}{2}\left
(\frac{\eta_0}{\xi_0}\right
)^2\frac{(1+\mu)^2}{\mu^2}+e^{-\psi(\alpha)}.
\ee
The value of $\psi(\alpha)$ can be obtained from Eq.
(\ref{eq.profildensite1dcc}). Then, Eqs. (\ref{eq.termeExi01dcc}), (\ref{cry})
and (\ref{mar2})  yield
\be
\label{celb}
E=(E-1)\frac{(1+\mu)^2}{\mu^2}+\frac{E}{\cosh^2\left\lbrack
\sqrt{\frac{E}{2}}\alpha(1-\zeta)+C\right\rbrack}.
\ee
Introducing the normalized inverse temperature
(\ref{etau}) and the normalized central density (\ref{chiw}) into Eq. (\ref{alpha})
we find that
\be
\label{fly}
\alpha=\sqrt{\eta\chi}.
\ee
We can therefore rewrite Eq. (\ref{celb}) as
\be
E=(E-1)\frac{(1+\mu)^2}{\mu^2}+\frac{E}{\cosh^2\left\lbrack
\sqrt{\frac{\eta\chi E}{2}}(1-\zeta)+C\right\rbrack}
\ee
or, equivalently, as
\be
\label{strong}
\frac{1+\mu}{\mu}\sqrt{\frac{E-1}{E}}=\tanh\left\lbrack
\sqrt{\frac{\eta\chi
E}{2}}(1-\zeta)+C\right\rbrack.
\ee
According to Eq. (\ref{lov}) we have
\be
\label{nigh}
\frac{\eta_0}{\xi_0}=\left (\frac{\eta}{\chi}\right )^{1/2}\mu.
\ee
Substituting (\ref{nigh}) into Eq. (\ref{eq.termeExi01dcc}) we obtain
\be
E=\frac{1}{2}\frac{\eta}{\chi}\mu^2+1.
\label{sag}
\ee
This equation determines $E$ as a function of $\chi$ (for given values of $\eta$ and $\mu$). Then 
Eq. (\ref{strong}) determines $\chi$ as a function of $\eta$ and $\mu$. 
Using Eqs. (\ref{rhopsi}), (\ref{mars1}), (\ref{cry}) and (\ref{fly})  we can write the density profile as
\be
\frac{\rho}{\rho_0}=\frac{E}{\cosh^2\left\lbrack\sqrt{\frac{\eta\chi E}{2}}\frac{x-x_*}{R}+C\right\rbrack}.
\ee
Then, using Eq. (\ref{mr}), we obtain the mass profile
\be
\frac{M(x)}{M}=\sqrt{\frac{2E\chi}{\eta}}\tanh\left (\sqrt{\frac{\eta\chi E}{2}}\frac{x-x_*}{R}+C\right )-\mu,
\ee
where we used Eqs. (\ref{saez}), (\ref{chiw}) and (\ref{sag}). We can easily
check that the foregoing equations are
equivalent to the equations of Sec. \ref{sec_elen}.

{\it Remark:} From Eqs. (\ref{etau}) and (\ref{xieta}) we find that
\be
\frac{\eta_0}{\eta}=\mu\zeta.
\ee
We then obtain
$\eta_0/\xi_0=\mu\zeta\eta/\alpha\zeta=\mu\eta/\alpha=\mu\eta/\sqrt{\chi\eta}
=\mu\sqrt{\eta/\chi}$ which returns Eq. (\ref{nigh}).


\section{Boltzmann-Poisson equation in two dimensions}
\label{app.boltzpoisson2d}

In $d=2$ dimensions, the Emden equation (\ref{emden}) reads
\be
\frac{1}{\xi}\frac{d}{d\xi}
\left(
\xi \frac{d \psi}{d\xi}
\right)
= e^{-\psi},
\label{emdend2}
\ee
where
\be
\xi=(2\pi\beta Gm\rho_0)^{1/2}r
\ee
is the rescaled distance. In the presence of a central body, this equation has to be solved with the boundary
conditions
\bea
\psi_{0} = 0 \quad {\rm and}\quad \psi'_{0} = \frac{\eta_0}{\xi_0}
\eea
at $\xi=\xi_0$, where $\xi_0$ and $\eta_0$ are defined by
\bea
\label{buff}
\xi_0=(2\pi\beta Gm\rho_0)^{1/2}R_*,\qquad \eta_0=\beta GmM_*.
\eea
We note the relation
\be
\label{mars6}
\xi=\xi_0\frac{r}{R_*}.
\ee
With the change of variables $t=\ln\xi$ and
$z=2\ln\xi-\psi$, the Emden equation can be reduced to the form
\be
\label{eq.equationmvtpartfictive}
\frac{d^{2}z}{dt^{2}}=-e^{z}.
\ee
This equation describes the motion of a fictive particle of unit mass in a
potential
$V(z)=e^{z}$ where $z$ plays the role of the position and $t$ the role of
the time. The particle starts at $t_{0}=\ln\xi_{0}$ from the position
$z_{0}=2\ln\xi_{0}$ with the velocity $(dz/dt)_{0}=2-\eta_{0}$. The energy
of the particle
\be
E=\frac{1}{2}\left (\frac{dz}{dt}\right )^{2}+e^{z}
\ee
is conserved. Using the initial condition at $t=t_{0}$, we find that
\be
E=\frac{1}{2}(2-\eta_{0})^{2}+\xi_{0}^{2}.
\ee
The solution of Eq. (\ref{eq.equationmvtpartfictive}) is
obtained by integrating the first order differential equation
\be
\label{eq.eqdiffboltzpoiss2d}
\frac{dz}{\sqrt{2(E-e^{z})}}=\pm dt.
\ee
We need to distinguish two cases depending on the sign of
$(dz/dt)_{0}=2-\eta_0$.


\subsection{The case $\eta_{0}<2$}
\label{sec_k1}

When $\eta_0<2$, the particle starts at $t=t_{0}$ from the position
$z_0=2\ln\xi_0$ with a
positive
velocity $(dz/dt)_{0}=2-\eta_0>0$, climbs the
potential $V(z)$ until the position $z_*=\ln E$  at which its velocity vanishes
(this happens at a time
$t_{*}$), and then
descends the potential. For $t_{0}<t<t_{*}$, we have $dz/dt>0$ (first
regime) and for $t>t_{*}$, we have $dz/dt<0$ (second
regime).


\subsubsection{First regime $t_{0}<t<t_{*}$}

For $t_{0}<t<t_{*}$, Eq. (\ref{eq.eqdiffboltzpoiss2d}) with the sign
$+$ can be integrated into
\be
\tanh^{-1}\sqrt{1-\frac{1}{E}e^{z}}=-\sqrt{\frac{E}{2}}t+C,
\ee
where $C$ is a constant.
Using the identity (\ref{id})
and the fact that $z=z_0=2\ln\xi_0$ at $t=t_{0}$, we find that the constant of
integration is given by
\be
C=\frac{1}{2}\ln\left
(\frac{1+\sqrt{1-\frac{1}{E}\xi_{0}^{2}}}{1-\sqrt{1-\frac{1}{E}\xi_{0}^{2}}}
\right )+\sqrt{\frac{E}{2}}\ln\xi_{0}.
\ee
Therefore, using Eq. (\ref{saez}), the solution of Eq. (\ref{eq.equationmvtpartfictive}) is
\be
e^{z}=\frac{E}{\cosh^{2}\left (-\sqrt{\frac{E}{2}}t+C\right )}.
\ee
This solution is valid for $t_{0}<t<t_{*}$. The time $t_{*}$ at which
the velocity of the particle vanishes is determined by the condition 
$e^{z_*}=E$ giving $t_{*}=C\sqrt{2/E}$.

Returning to the original variables and setting $F=\sqrt{2E}$ and
$\lambda^{2}=e^{-2C}$, we obtain 
\be
\label{eq.epsi2dt<t*}
e^{-\psi}=\frac{2F^{2}\lambda^{2}\xi^{F-2}}{\left (1+\lambda^{2}\xi^{F}\right )^{2}}
\ee
with
\be
\label{eq.defF2d}
F=\sqrt{(2-\eta_{0})^{2}+2\xi_{0}^{2}}
\ee
and
\be
\label{eq.lambda2eta0<2}
\lambda^{2}=\frac{1}{\xi_{0}^{F}}
\frac{1-\sqrt{1-\frac{2\xi_{0}^{2}}{F^{2}}}}{1+\sqrt{1-\frac{2\xi_{0}^{2}}{F^{2}
}}}.
\ee
This solution is valid for $\xi_0\le \xi\le \xi_*$ with $\xi_*=1/\lambda^{2/F}$.


\subsubsection{Second regime $t>t_{*}$}

For $t>t_{*}$, Eq. (\ref{eq.eqdiffboltzpoiss2d}) with the sign $-$ can
be integrated into
\be
\tanh^{-1}\sqrt{1-\frac{1}{E}e^{z}}=\sqrt{\frac{E}{2}}t+D,
\ee
where $D$ is a constant.
Therefore, the solution of Eq. (\ref{eq.equationmvtpartfictive}) is
\be
e^{z}=\frac{E}{\cosh^{2}\left (\sqrt{\frac{E}{2}}t+D\right )}.
\ee
The matching condition at $t=t_{*}$ where $e^{z_*}=E$ gives
$t_{*}=-D\sqrt{2/E}$. Comparing this expression with the expression found in
the previous section, we conclude that $D=-C$. This implies that the solution
(\ref{eq.epsi2dt<t*})-(\ref{eq.lambda2eta0<2}) is actually valid for
all
$\xi\ge\xi_{0}$. For $\xi\rightarrow +\infty$, we have
\be
e^{-\psi}\sim \frac{2F^2}{\lambda^2}\frac{1}{\xi^{F+2}}.
\label{asy}
\ee
For $\xi\rightarrow \xi_0$, we have
\be
e^{-\psi}\simeq 1-\frac{\eta_0}{\xi_0}(\xi-\xi_0)+...,
\ee
in agreement with Eq. (40) of \cite{css}. The gas forms a
central cusp of typical size $\xi_0/\eta_0$ at the contact with the central body.
Without the central body
($\eta_{0}=\xi_{0}=0$), we find that $F=2$ and $\lambda^{2}=1/8$, returning the
Ostriker solution \cite{ostriker,sirechavanis2002,chavanis2007}
\be
e^{-\psi}=\frac{1}{(1+\frac{1}{8}\xi^{2})^{2}}.
\ee
%


\subsection{The case $\eta_{0}>2$}
\label{sec_k2}
 
When $\eta_0>2$, the particle starts at $t=t_{0}$ from the  position
$z_0=2\ln\xi_0$ with a
negative velocity $(dz/dt)_{0}=2-\eta_0<0$ and descends
the potential. For $t>t_{0}$, we have $dz/dt<0$. Eq.
(\ref{eq.eqdiffboltzpoiss2d}) with the sign $-$ can be integrated
into
\be
\tanh^{-1}\sqrt{1-\frac{1}{E}e^{z}}=\sqrt{\frac{E}{2}}t+D,
\ee
where $D$ is a constant.
Using the identity (\ref{id}) and the fact that 
$z=z_0=2\ln\xi_0$ at
$t=t_{0}$, we find that the constant of integration is given by
\be
D=\frac{1}{2}\ln\left
(\frac{1+\sqrt{1-\xi_{0}^{2}/E}}{1-\sqrt{1-\xi_{0}^{2}/E}}\right
)-\sqrt{\frac{E}{2}}\ln\xi_{0}.
\ee
Therefore, the solution of Eq. (\ref{eq.equationmvtpartfictive}) is
\be
e^{z}=\frac{E}{\cosh^{2}\left (\sqrt{\frac{E}{2}}t+D\right )}.
\ee
Returning to the original variables and setting $F=\sqrt{2E}$ and
$\lambda^{2}=e^{2D}$, we obtain
\be
\label{autre}
e^{-\psi}=\frac{2F^{2}\lambda^{2}\xi^{F-2}}{\left (1+\lambda^{2}\xi^{F}\right )^{2}}
\ee
with
\be
F=\sqrt{(2-\eta_{0})^{2}+2\xi_{0}^{2}}
\label{maef}
\ee
and
\be
\label{eq.lambda2eta0>2}
\lambda^{2}=\frac{1}{\xi_{0}^{F}}
\frac{1+\sqrt{1-\frac{2\xi_{0}^{2}}{F^{2}}}}{1-\sqrt{1-\frac{2\xi_{0}^{2}}{F^{2}
}}}.
\ee
We note that the signs $\pm$ are reversed in the expression of $\lambda^2$ given
by Eq. (\ref{eq.lambda2eta0>2})  as compared to Eq. (\ref{eq.lambda2eta0<2}).
However, using Eq. (\ref{maef}) and
paying a careful attention to the sign of $2-\eta_0$ we can easily show that
Eqs. (\ref{eq.lambda2eta0<2}) and (\ref{eq.lambda2eta0>2}) can both be rewritten
as 
\be
\label{mieuF}
\lambda^2\xi_0^F=\frac{F+\mu\eta-2}{F-\mu\eta+2},
\ee
so they coincide. Therefore, the results of Appendices \ref{sec_k1} and \ref{sec_k2} are equivalent.


We now have to determine the central density $\rho_0$ as a function of
the temperature $T$. To that purpose, we will use Newton's law which can
be written in
$d=2$ as [see Eq. (\ref{mxi})]
\be
\frac{d\psi}{d\xi}=\eta_0\left\lbrack
1+\frac{M(\xi)}{M_*}\right\rbrack \frac{1}{\xi}.
\label{gfbw}
\ee
Using Eqs. (\ref{mu}), (\ref{dec}),
(\ref{chh}) and (\ref{buff}) we find that
\be
\label{eq.xi0eta0}
\eta_{0}=\mu\eta,\qquad \xi_{0}^{2}=\chi\eta.
\ee
Therefore, Eq. (\ref{gfbw}) can be rewritten as
\be
\label{dit}
\xi\psi'(\xi)=\eta \left (\mu+\frac{M(\xi)}{M}\right ).
\ee

\subsection{Infinite domain}

In an infinite domain, taking the limit $\xi\rightarrow +\infty$ in Eq.
(\ref{dit}), we
obtain
\be
\label{eq.limitexipsi2d}
\lim_{\xi\rightarrow +\infty} \xi\psi'(\xi)=\eta (\mu+1).
\ee
From Eq. (\ref{asy}), we find that
\be
\psi'\sim \frac{F+2}{\xi}, \qquad (\xi\rightarrow +\infty).
\ee
Therefore
\be
\label{eq.F2d}
F=\eta(1+\mu)-2.
\ee
Substituting Eq. (\ref{eq.xi0eta0}) into  Eq. (\ref{eq.defF2d}), we get
\be
\label{artg}
F=\sqrt{(2-\mu\eta)^2+2\chi\eta},
\ee
which coincides with Eq. (\ref{fd}). Then Eq. (\ref{eq.F2d}) returns Eq.
(\ref{eq.Finf2d}). 
From Eqs. (\ref{eq.F2d}) and
(\ref{artg}), we
find that
\be
\label{terre}
\chi=\frac{2\mu+1}{2}\eta-2,
\ee
which returns Eq. (\ref{eq.chieta}). This relation determines the
normalized central density $\chi$ as a function of the normalized inverse
temperature $\eta$ (for a given value of $\mu$).

According to Eqs. (\ref{rhopsi}) and (\ref{eq.epsi2dt<t*}) the
density profile is given by
\be
\frac{\rho(r)}{\rho_0}=\frac{2F^{2}\lambda^{2}\xi^{F-2}}{\left
(1+\lambda^{2}\xi^{F}\right )^{2}}.
\ee
Using Eq. (\ref{mars6}), it
can be rewritten as
\be
\label{pan2}
\rho(r)=\rho_{0}\frac{2F^{2}\lambda^{2}\xi_{0}^{F-2}(r/R_{*})^{F-2}}{
\left\lbrack 1+\lambda^{2}\xi_{0}^{F}(r/R_{*})^{F}\right \rbrack^{2}}.
\ee
Using (\ref{eq.defF2d}),
(\ref{eq.lambda2eta0<2}) and (\ref{eq.xi0eta0}), we find that
\be
\lambda^{2}\xi_{0}^{F}=\frac{F-2+\mu\eta}{F+2-\mu\eta}.
\ee
Then, using Eqs. (\ref{eq.F2d}) and (\ref{terre}), we get
\be
\label{pan}
\lambda^{2}\xi_{0}^{F}=\frac{(2\mu+1)\eta-4}{\eta}=\frac{2\chi}{\eta}.
\ee
According to Eqs. (\ref{eq.termeK2d}) and (\ref{pan}), we have
\be
\lambda^{2}\xi_{0}^{F}=K,
\ee
where $K$ is defined by Eq.
(\ref{kd}). Finally, we find that
$2F^{2}\lambda^{2}\xi_{0}^{F-2}=2F^2K/\xi_0^2=2F^2K/\chi\eta=4F^{2}/\eta^{2}
=(1+K)^{2}$ where we have used Eq. (\ref{eq.termeK2d}). Therefore,
we can rewrite the density profile (\ref{pan2}) as
\be
\rho=\rho_{0}\frac{(1+K)^{2}(r/R_{*})^{F-2}}{\left\lbrack 1+K(r/R_{*})^{F}\right\rbrack^{2}},
\ee
which returns Eq. (\ref{eq.rhoextinf2dcc}). By integration, we
obtain the mass profile (\ref{eq.masseinf2d}). In conclusion, we recover the
results of Sec. \ref{sec_lum1}.

\subsection{Finite domain}

In a finite domain, applying Eq. (\ref{dit}) on the boundary, we obtain
\be
\label{route}
\alpha\psi'(\alpha)=\eta(\mu+1),
\ee
where
\be
\label{hamo}
\alpha=(2\pi\beta Gm\rho_0)^{1/2}R
\ee
is the normalized box radius.  Computing $\psi'(\alpha)$ from Eq.
(\ref{eq.epsi2dt<t*}) and substituting
its value  into Eq. (\ref{route}) we obtain
\be
\label{fort}
\frac{2\lambda^2F\alpha^F}{1+\lambda^2\alpha^F}-(F-2)=\eta(\mu+1).
\ee
According to Eqs. (\ref{buff}), (\ref{eq.xi0eta0}) and (\ref{hamo}), we have
\be
\label{aime}
\alpha=\frac{\xi_0}{\zeta}=\frac{\sqrt{\eta\chi}}{\zeta},
\ee
where $\zeta$ is defined by Eq. (\ref{mu}). 
Using Eq. (\ref{eq.xi0eta0}), we can express $F$ and $\lambda$ 
[defined in Eqs. (\ref{eq.defF2d}) and (\ref{eq.lambda2eta0<2})] as a function
of $\chi$ (for given $\eta$ and $\mu$). On the other hand, Eq. (\ref{aime})
gives $\alpha$ as a function of $\chi$ (for given $\eta$ and $\zeta$).
Therefore, Eq. (\ref{fort})
determines the dimensionless central density $\chi$ as a function of
the normalized inverse temperature $\eta$ for given values of $\mu$ and $\zeta$. We now show the equivalence with the
results of Sec. \ref{sec_lum2}.

Substituting Eq. (\ref{eq.xi0eta0}) into Eq. (\ref{eq.defF2d})
we get
\be
\label{esp}
F=\sqrt{(2-\mu\eta)^2+2\chi\eta},
\ee
which coincides with Eq. (\ref{fd}). On the other hand, using Eq. (\ref{aime})
we can rewrite Eq. (\ref{fort})
as
\be
\label{pole}
\frac{2\lambda^2F\frac{\xi_0^F}{\zeta^F}}{1+\lambda^2\frac{\xi_0^F}{\zeta^F}}
=\eta(\mu+1)+F-2.
\ee
Using Eqs. (\ref{eq.defF2d}), (\ref{eq.lambda2eta0<2}) and (\ref{eq.xi0eta0}) we get  
\be
\label{pense}
\lambda^{2}\xi_{0}^{F}=\frac{F-2+\mu\eta}{F+2-\mu\eta}=\frac{K}{M_1},
\ee
where $M_1$ and $K$ are defined by Eqs. (\ref{m1d}) and (\ref{kd}). With this relation
we can rewrite Eq. (\ref{pole}) as
\be
\frac{2F}{1+\zeta^F \frac{M_1}{K}}=\eta(\mu+1)+F-2
\ee
or, equivalently, as
\be
\zeta^F=\frac{K}{M_1}\frac{F+2-\eta(1+\mu)}{F-2+\eta(1+\mu)}.
\ee
Using Eq. (\ref{cime}), it is easy to show
that this equation coincides with Eq. (\ref{jaz}). Therefore, Eq. (\ref{fort}) is equivalent to Eq. (\ref{jaz}).

On the other hand, according to Eqs. (\ref{rhopsi}) and  (\ref{eq.epsi2dt<t*}) the density profile is given
by
\be
\frac{\rho(r)}{\rho_0}=\frac{2F^{2}\lambda^{2}\xi^{F-2}}{\left
(1+\lambda^{2}\xi^{F}\right )^{2}}.
\ee
Using Eq. (\ref{mars6}) it can be rewritten as
\be
\rho(r)=\rho_{0}\frac{2F^{2}\lambda^{2}\xi_{0}^{F-2}(r/R_{*})^{F-2}}{
\left\lbrack
1+\lambda^{2}\xi_{0}^{F}(r/R_{*})^{F}\right \rbrack^{2}}.
\ee
With the relation from Eq. (\ref{pense}) we obtain
\be
\label{mahe}
\rho(r)=\rho_{0}\frac{2F^{2}\frac{K}{M_1}\frac{1}{\xi_0^2}
(r/R_{*})^{F-2}}{\left\lbrack
1+\frac{K}{M_1}(r/R_{*})^{F}\right \rbrack^{2}}.
\ee
In order to show the equivalence between Eq. (\ref{mahe}) and Eq. (\ref{bruit})
we have to establish that
\be
\label{mir}
2F^2\frac{K}{M_1}\frac{1}{\xi_0^2}=\left (1+\frac{K}{M_1}\right )^2.
\ee
Since this equality is not straightforward, we detail the different
steps below. Using Eq. (\ref{eq.xi0eta0}), we find that Eq. (\ref{mir}) is equivalent
to
\be
\label{art}
\frac{2F^2 K}{M_1\eta\chi}=\left (1+\frac{K}{M_1}\right )^2.
\ee
Expanding the right hand side of Eq. (\ref{art}) we get
\be
\left (\frac{K}{M_1}\right )^2+2\left
(1-\frac{F^2}{\eta\chi}\right )\frac{K}{M_1}+1=0.
\ee
Solving this second degree equation we find that Eq. (\ref{mir}) is equivalent
to
\be
\frac{K}{M_1}=-1+\frac{F^2}{\eta\chi}\pm
\frac{F^2}{\eta\chi}\sqrt{1-\frac{2\eta\chi}{F^2}}.
\label{rel}
\ee
Now, according to Eq. (\ref{esp}), we have
\be
1-\frac{2\eta\chi}{F^2}=\frac{(2-\eta\mu)^2}{F^2}.
\ee
Therefore, Eq. (\ref{rel}) is equivalent to
\be
\frac{K}{M_1}=-1+\frac{F^2}{\eta\chi}\pm
\frac{F}{\eta\chi}(2-\mu\eta)
\ee
or, using Eq. (\ref{esp}), to
\be
\label{cri}
\frac{K}{M_1}=1+\frac{(2-\eta\mu)^2}{\eta\chi}\pm
\frac{F}{\eta\chi}(2-\mu\eta).
\ee
According to Eq. (\ref{cime}), we have
\be
\label{ber}
\frac{K}{M_1}-1=\frac{2(\eta\mu-2)}{2-\eta\mu+F}.
\ee
In order to establish Eq. (\ref{cri}), hence Eq. (\ref{mir}), we thus have to show that
\be
\frac{2(\eta\mu-2)}{2-\eta\mu+F}=\frac{(2-\eta\mu)(2-\eta\mu\pm F)}{\eta\chi}
\ee
or, equivalently, that
\be
2\eta\chi=-(2-\eta\mu+F)(2-\eta\mu\pm F).
\ee
We see that this condition with the sign $-$ is equivalent to Eq.
(\ref{esp}).
Therefore, we have shown that Eq. (\ref{mir}) holds so that Eq. (\ref{mahe})
returns Eq. (\ref{bruit}). By integration, we the obtain Eq. (\ref{pleur}).
In conclusion, we recover the results of Sec. \ref{sec_lum2}.

\subsection{The case of a central Dirac mass}

In the presence of a central Dirac mass, the Boltzmann-Poisson equation is
given by Eq. (\ref{eq.equationboltzpoiss}). Introducing the dimensionless
variables $\psi=\beta m\Phi_{\rm tot}$ and $\xi=(S_d \beta G m A)^{1/2}r$ we can
write the density profile as 
\be
\label{era1}
\rho=Ae^{-\psi},
\ee
where $\psi$ satisfies the generalized Emden equation
\be
\label{era2}
\frac{1}{\xi^{d-1}}\frac{d}{d\xi}\left (\xi^{d-1}\frac{d\psi}{d\xi}\right )=e^{-\psi}+\frac{M_*}{A}\frac{\delta(\xi)}{S_d\xi^{d-1}}.
\ee
We now focus on the dimension $d=2$. Integrating Eq. (\ref{dain2}) we 
find that $\Phi_{\rm tot}\sim GM_*\ln r+{\rm cst}$ for $r\rightarrow 0$. This
gives $\psi\sim \eta_0\ln\xi+{\rm cst}$ for $\xi\rightarrow 0$. When $\xi>0$,
Eq. (\ref{era2}) reduces to Eq. (\ref{emdend2}). Repeating the calculations
presented at the beginning of this Appendix, we obtain Eq. (\ref{eq.epsi2dt<t*})
with $F=2-\eta_0>0$. Since $\eta_0=\mu\eta$, we get
\be
\label{era3}
F=2-\mu\eta.
\ee
On the other hand, Eq. (\ref{dit}) remains
valid.

{\it Infinite domain:} Using Eqs. (\ref{eq.F2d}) and (\ref{era3}) we find that
\be
\label{era4}
\eta=\frac{4}{1+2\mu}.
\ee
The equilibrium states exist at a unique temperature. This returns Eq.
(\ref{eq.etainf2d}). The density profile is given by Eqs. (\ref{era1}) and
(\ref{eq.epsi2dt<t*}). Writing $\xi=kr$ with $k=(2\pi\beta GmA)^{1/2}$, defining
$Q=1/(\lambda^2k^F)$, and using $F=\beta G M m/2$, we recover Eq.
(\ref{surprise}).

{\it Finite domain:}  Eq. (\ref{route}) remains valid with $\alpha=(2\pi\beta
GmA)^{1/2}R=kR$. We then obtain Eq. (\ref{fort}) with $F$ given by Eq.
(\ref{era3}). The density profile is given by Eqs. (\ref{era1}) and
(\ref{eq.epsi2dt<t*}). Writing $\xi=r\alpha/R$, using $F=M_1\eta/2$ where $M_1$
is defined by Eq. (\ref{m1v}), and using the relation
$1/(\lambda^2\alpha^F)=M_1-1$ 
deduced from Eq. (\ref{fort}), we recover Eq. (\ref{eq.rhoanalytique2dfini}).

\section{The potential in $d=2$ dimensions for an axisymmetric system}

For an axisymmetric system, the gravitational potential in $d=2$ dimensions can be written as (see Appendix B of \cite{css})
\be
\label{fin1}
\Phi(r)=G\int_{R_*}^{R}\int_{0}^{2\pi} \rho(r_1)\ln \frac{|{\bf r}-{\bf r}_1|}{R}\, r_1 dr_1 d\theta_1.
\ee
Introducing the expansion
\be
\label{fin2}
\ln |{\bf r}-{\bf r}_1|=\ln r_>-\frac{1}{2}\sum_{n\neq 0}\frac{1}{|n|}\left (\frac{r_<}{r_>}\right )^{|n|}e^{i n\phi},
\ee
where $r_<={\rm min}(r,r_1)$, $r_>={\rm max}(r,r_1)$ and $\phi=\theta-\theta_1$ in Eq. (\ref{fin1}), we find that
\begin{eqnarray}
\label{fin3}
\Phi(r)=2\pi G \ln \left (\frac{r}{R}\right ) \int_{R_*}^{r} \rho(r_1) r_1\, dr_1\nonumber\\
+2\pi G \int_r^{R} \rho(r_1)\ln \left (\frac{r_1}{R}\right ) r_1\, dr_1.
\end{eqnarray}
From this expression, we get
\begin{eqnarray}
\label{fin4}
\Phi(R_*)=2\pi G \int_{R_*}^{R} \rho(r)\ln \left (\frac{r}{R}\right ) r\, dr.
\end{eqnarray}

The gravitational potential in $d=1$ dimension can be written as (see Appendix B of \cite{css})
\be
\label{fin5}
\Phi(x)=G\int_{x_*}^{R}\rho(x_1)|x-x_1|\, dx_1
\ee
or, equivalently, as
\begin{equation}
\label{fin3}
\Phi(x)=G\int_{x_*}^{x}\rho(x_1)(x-x_1)\, dx_1+G\int_{x}^{R}\rho(x_1)(x_1-x)\, dx_1.
\end{equation}
From this expression, we get
\begin{eqnarray}
\label{fin4}
\Phi(x_*)=G\int_{x_*}^{R}\rho(x)(x-x_*)\, dx.
\end{eqnarray}

\end{appendix}



\begin{thebibliography}{99}





\bibitem{emden}  {\small R. Emden, {\it Gaskugeln}  (Leipzig, 1907)}
\bibitem{chandrass} {\small S. Chandrasekhar, {\it An Introduction to the
Theory of Stellar Structure} (Dover, New York, 1939)}
\bibitem{paddy} {\small T. Padmanabhan, Phys. Rep.  {\bf 188}, 285 (1990)}
\bibitem{dvs1}  {\small H.J. de Vega, N. Sanchez, Nucl. Phys. B  {\bf
625}, 409 (2002)}
\bibitem{dvs2}  {\small H.J. de Vega, N. Sanchez,
Nucl. Phys. B  {\bf 625}, 460 (2002)}
\bibitem{found}  {\small J. Katz,  Found. Phys. {\bf 33}, 223 (2003)}
\bibitem{ijmpb} {\small P.H. Chavanis, Int. J. Mod. Phys. B {\bf 20}, 3113 (2006)}
\bibitem{amba}{\small V.A. Ambarzumian, Ann. Leningrad State University {\bf
22}, 19 (1938)}
\bibitem{spitzer1940}{\small L. Spitzer, Mon. Not. R. Astron. Soc. {\bf 100},
396 (1940)}
\bibitem{chandra}  {\small S. Chandrasekhar, {\it Principles of Stellar Dynamics} (University of Chicago press, 1942)}
\bibitem{michie} {\small R.W. Michie,  Mon. Not. R. Astron. Soc. {\bf 125}, 127
(1963)}
\bibitem{king} {\small I.R. King, Astron. J. {\bf 70}, 376 (1965)}
\bibitem{antonov} {\small V.A. Antonov, Vest. Leningr. Gos. Univ. {\bf 7}, 135
(1962).}
\bibitem{lbw} {\small D. Lynden-Bell, R. Wood, Mon. Not. R. Astron. Soc. {\bf
138}, 495 (1968).}
\bibitem{katzking} {\small J. Katz,  Mon. Not. R. Astron. Soc. {\bf 190}, 497
(1980)}
\bibitem{clm1}{\small P.H. Chavanis, M. Lemou, F. M\'ehats,  Phys. Rev. D
{\bf 91}, 063531 (2015)}
\bibitem{larson} {\small R.B. Larson,   Mon. Not. R. Astron.
Soc. {\bf 147}, 323 (1970)}
\bibitem{hachisu} {\small  I. Hachisu, Y. Nakada, K. Nomoto, D. Sugimoto,
Prog. Theor. Phys.  {\bf 60}, 393 (1978)}
\bibitem{lbe} {\small D. Lynden-Bell, P.P. Eggleton,  Mon. Not. R.
Astron. Soc. {\bf 191}, 483 (1980)}
\bibitem{cohn} {\small H. Cohn, Astrophys. J. {\bf 242}, 765 (1980)}
\bibitem{henonbinary}  {\small M. H\'enon, Ann. Astrophys. {\bf 24}, 369
(1961)}
\bibitem{inagakilb} {\small S. Inagaki, D. Lynden-Bell,  Mon. Not. R. Astron.
Soc. {\bf 205}, 913 (1983)}
\bibitem{hs} {\small D.C. Heggie, D. Stevenson,  Mon. Not. R. Astron.
Soc. {\bf 230}, 223 (1988)}
\bibitem{sugimoto} {\small D. Sugimoto, E. Bettwieser,   Mon. Not. R. Astron.
Soc. {\bf 204}, 19 (1983)}
\bibitem{camm}  {\small G.L. Camm, Mon. Not. R. Astron. Soc.   {\bf 110}, 305 (1950)}
\bibitem{ostriker}  {\small J. Ostriker,  ApJ {\bf 140}, 1056 (1964)}
\bibitem{css}  {\small P.H. Chavanis, J. Sopik, C. Sire,  Phys. Rev. E {\bf 109}, 014118 (2024)}
\bibitem{spitzer}  {\small L. Spitzer, Astrophys. J.  {\bf 95}, 329 (1942)}
\bibitem{ledoux}  {\small P. Ledoux, Ann. Astrophys.  {\bf 14}, 438 (1951)}
\bibitem{pacholczyk}  {\small A.G. Pacholczyk, Acta Astr. {\bf 13}, 1 (1963)}
\bibitem{stodolkiewicz}  {\small J.S. Stodolkiewicz, Acta Astr. {\bf 13}, 30 (1963)}
\bibitem{rybicki}  {\small G.B. Rybicki,  Astr. Space. Sci. {\bf 14}, 56 (1971)}
\bibitem{hlake}  {\small E.R. Harrison, R.G. Lake, Astrophys. J.  {\bf 171}, 323 (1972)}
\bibitem{sirechavanis2002}{\small  C. Sire, P.H. Chavanis, Phys. Rev. E {\bf 66}, 046133 (2002)}
\bibitem{chavanissire2006a}{\small  P.H. Chavanis, C. Sire, Phys. Rev. E {\bf 73}, 066103 (2006)}
\bibitem{chavanis2007}  {\small P.H. Chavanis, Physica A  {\bf 384}, 392 (2007)}
\bibitem{muller}{\small B. Bakhti, D. Boukari, M. Karbach, P. Maass, and G. M\"uller, Phys. Rev. E {\bf 97}, 042131 (2018)}
\bibitem{katzlecar}  {\small J. Katz, M. Lecar,  Astr. Space. Sci. {\bf 68}, 495 (1980)}
\bibitem{chavanis2006}  {\small P.H. Chavanis, C. R. Physique  {\bf 7}, 331 (2006)}
\bibitem{salzberg}  {\small A.M. Salzberg, J. Math. Phys.  {\bf 6}, 158 (1965)}
\bibitem{aakin}{\small P.H. Chavanis, Astron. Astrophys. {\bf 556}, A93 (2013)} 
\bibitem{chavanissire2006b}{\small  P.H. Chavanis, C. Sire, Phys. Rev. E {\bf
73}, 066104 (2006)}
\bibitem{chavanis2012}  {\small P.H. Chavanis, Int. J. Mod. Phys. B {\bf 26}, 1241002 (2012)}
\bibitem{chavanisDexact}{\small  P.H. Chavanis, Eur. Phys. J. B {\bf 57}, 391 (2007) }
\bibitem{chavanismannella}{\small  P.H. Chavanis, R. Mannella, Eur. Phys. J. B {\bf 78}, 139 (2010)}
\bibitem{poincare}  {\small H. Poincar\'e, Acta Math. {\bf 7}, 259 (1885)}
\bibitem{katzpoincare1}  {\small J. Katz, Mon. Not. R. Astron. Soc. {\bf 183},
765 (1978)}
\bibitem{acepjb}  {\small G. Alberti, P.H. Chavanis, Eur. Phys. J. B {\bf 93},
208 (2020)}
\bibitem{ptd}  {\small P.H. Chavanis, Phys. Rev. E {\bf 69}, 066126 (2004)}
\bibitem{williamson}  {\small  J.H. Williamson, J. Plasma Physics {\bf  17}, 85
(1977)}
\bibitem{caglioti}  {\small E. Caglioti, P.L. Lions, C. Marchioro, M. 
Pulvirenti,  Commun. Math. Phys.   {\bf 143}, 501 (1992)}
\bibitem{kiessling}  {\small  M. Kiessling, J. Plasma Physics  {\bf 54}, 11
(1995)}
\bibitem{houchesPH}  {\small P.H. Chavanis,  {Statistical mechanics of
two-dimensional vortices and stellar systems}, in:  Dynamics and thermodynamics
of systems with long range interactions, edited by Dauxois, T, Ruffo, S.,
Arimondo, E. and  Wilkens, M. Lecture Notes in Physics, Springer (2002)}
\bibitem{virialonsager}  {\small  P.H. Chavanis, Eur. Phys. J. Plus  {\bf 127},
159 (2012)}
\bibitem{lp}  {\small T.S. Lundgren, Y.B. Pointin, J. Stat. Phys. {\bf 17}, 323
(1977)}
\bibitem{bennett}  {\small W.H. Bennett, Phys. Rev.   {\bf 45}, 890 (1934)}
\bibitem{kl2}  {\small M. Kiessling, J. Lebowitz, Phys. Plasmas {\bf 1}, 1841
(1994)}
\bibitem{cp}  {\small S. Childress, J.K. Percus, Math. Biosci. {\bf 56}, 217
(1981)}
\bibitem{hv}  {\small M.A. Herrero, J.J.L. Velazquez, J. Math. Biol. {\bf 35},
177 (1996)}
\bibitem{bkln}  {\small P. Biler, G. Karch, P. Lauren\c cot, T. Nadzieja, Math.
Methods Appl. Sci. {\bf 29}, 1563 (2006)}
\bibitem{gervaisneveu}  {\small J.L. Gervais, A. Neveu, Nucl. Phys. B {\bf 199},
59 (1982)}
\bibitem{liouville}  {\small J. Liouville, J. Math. Pures Appl. {\bf 18}, 71
(1853)}
\bibitem{alyperez}  {\small J.J Aly, J. Perez, Phys. Rev. E {\bf 60}, 5185
(1999)}
\bibitem{katzlyndenbell}  {\small J. Katz, D. Lynden-Bell, Mon. Not. R. Astron. Soc. {\bf 184}, 709 (1978)}
\bibitem{padmanabhan91}  {\small T. Padmanabhan, Mon. Not. R. Astron. Soc.   {\bf 253}, 445 (1991)}
\bibitem{aly}  {\small J.J Aly, Phys. Rev. E {\bf 49}, 3771 (1994)}
\bibitem{abdallareza}  {\small E. Abdalla, M. Reza Rahimi Tabar, Phys. Lett. B {\bf 440}, 339 (1998)}
\bibitem{chandrafermi}  {\small S. Chandrasekhar, E. Fermi, Astrophys. J.  {\bf 118}, 116 (1953)}
\bibitem{kmcfermions}  {\small  M. Kirejczyk, G. M\"uller, P.H.
Chavanis, Phys. Rev. E {\bf 106}, 024132 (2022)}
\bibitem{ribot}  {\small P.H. Chavanis, M. Ribot, C. Rosier, C.  Sire, Banach
Center Publ. {\bf 66}, 103 (2004)}
\bibitem{these}  {\small  P.H. Chavanis, Contribution \`a la M\'ecanique
Statistique des Tourbillons Bidimensionnels. Analogie avec la Relaxation
Violente des Syst\`emes Stellaires. PhD thesis (ENS Lyon, 1996) }
\bibitem{tcfd}  {\small P.H. Chavanis, Theor. Comput. Fluid Dyn. {\bf 24}, 217
(2010)}
\bibitem{unified}{\small P.H. Chavanis, J. Stat. Mech. {\bf 5}, 05019 (2010)}
\bibitem{degrad}  {\small P.H. Chavanis,  Eur. Phys. J. B {\bf 54}, 525 (2006)}
\bibitem{crs}{\small  P.H. Chavanis, C. Rosier, C. Sire,  Phys. Rev. E {\bf 66}, 036105 (2002)}
\bibitem{post}{\small  C. Sire, P.H. Chavanis,  Phys. Rev. E {\bf 69}, 066109 (2004)}
\bibitem{nfp}{\small  P.H. Chavanis, Eur. Phys. J. B {\bf 62}, 179 (2008)}
\bibitem{ggpp}{\small P.H. Chavanis, Eur. Phys. J. Plus {\bf 132}, 248 (2017)}
\bibitem{ggppBH}{\small P.H. Chavanis, Eur. Phys. J. Plus {\bf 134}, 352 (2019)}
\bibitem{aaiso}  {\small P.H. Chavanis,  Astron. Astrophys. {\bf 381}, 340
(2002)}
\bibitem{estimate}{\small  P.H. Chavanis, C. Sire,  Phys. Rev. E {\bf 70}, 026115 (2004)}
\bibitem{Tzero}{\small  P.H. Chavanis, C. Sire,  Phys. Rev. E {\bf 83}, 031131 (2011)}
\bibitem{klimontovich}  {\small Y.L. Klimontovich, {\it The Statistical Theory
of Non-Equilibrium Processes in a Plasma} (M.I.T. press, Cambridge, 1967)}
\bibitem{vlasov1}  {\small A.A. Vlasov, Zh. Eksp. i Teor. Fiz.  {\bf 8}, 291
(1938)}
\bibitem{vlasov2}  {\small  A.A. Vlasov, J. Phys. (U.S.S.R.) {\bf  9}, 25
(1945)}
\bibitem{kirchhoff}{\small G. Kirchhoff, in {\it Lectures in Mathematical
Physics, Mechanics} (Teubner, Leipzig,
1877)}
\bibitem{onsager}  {\small L. Onsager, Nuovo Cimento Suppl. {\bf 6}, 279 (1949)}
\bibitem{jm}  {\small G. Joyce, D. Montgomery, J. Plasma Phys. {\bf 10}, 107 (1973)}
\bibitem{pl}  {\small Y.B. Pointin, T.S. Lundgren, Phys. Fluids. {\bf 19}, 1459 (1976)}
\bibitem{bv}  {\small  P.H. Chavanis,  Physica A {\bf 387}, 6917 (2008)}
\bibitem{Kvortex2023}{\small P.H. Chavanis, Eur. Phys. J. Plus {\bf 138},
136 (2023)}






\end{thebibliography}


\end{document}